%% file: main.tex
\documentclass[preprint2]{aastex6}

\usepackage{amsmath}
\usepackage{rotating}

\newcommand{\ExB}{\ensuremath{{\bf E}\times{\bf B}}}
\newcommand{\GHz}{\text{GHz}}
\newcommand{\kHz}{\text{kHz}}
\newcommand{\MHz}{\text{MHz}}

\newcommand{\ms}{\text{ms}}
\newcommand{\phase}{\ensuremath{\varphi}}
\newcommand{\pulsar}{PSR J0034$-$0721}
\newcommand{\pulsartwo}{PSR B0826$-$34}
\newcommand{\seconds}{\text{s}}
\newcommand{\tvar}{\ensuremath{p}}
\newcommand{\us}{\mu\seconds}

\newcommand{\dv}[2]{\frac{\text{d}{#1}}{\text{d}{#2}}}

\newcommand{\GG}{GGKS04}

\newcommand{\phaselo}{-25.3}
\newcommand{\phasehi}{33.0}
\newcommand{\convgausswidth}{2}
\newcommand{\minseparation}{10}
\newcommand{\drestimate}{-2.5}
\newcommand{\shortseqmaxlength}{4}
\newcommand{\pslength}{101}
\newcommand{\pssaturationpercent}{25}
\newcommand{\sigmathresh}{4}
\newcommand{\offpulsebuffer}{90}

\begin{document}

\title{Low Frequency Observations of the Subpulse Drifter \pulsar{} with the Murchison Widefield Array}

\author{S. J. McSweeney\altaffilmark{1} and N. D. R. Bhat\altaffilmark{1} and S. E. Tremblay\altaffilmark{1}}
\affil{International Centre for Radio Astronomy Research (ICRAR), Curtin University \\
1 Turner Ave., Technology Park, \\
Bentley, 6102, W.A., Australia}

\author{A. A. Deshpande}
\affil{Raman Research Institute (RRI) \\
C. V. Raman Avenue, \\
Sadashivanagar, \\
Bengaluru - 560 080, India}

\and

\author{S. M. Ord}
\affil{Commonwealth Scientific and Industrial Research Organisation (CSIRO) \\
Corner Vimiera \& Pembroke Roads \\
Marsfield NSW 2122 \\
Australia}

\altaffiltext{1}{ARC Centre of Excellence for All-Sky Astronomy (CAASTRO)}

\begin{abstract}

The phenomenon of subpulse drifting may hold the key to understanding the pulsar emission mechanism.
Here, we report on new observations of \pulsar{} (B0031$-$07), carried out with the Murchison Widefield Array at $185\,\MHz$.
We observe three distinct drift modes whose ``vertical'' drift band separations ($P_3$) and relative abundances are consistent with previous studies at similar and higher frequencies.
The driftbands, however, are observed to change their slopes over the course of individual drift modes, which can be interpreted as a continuously changing drift rate.
The implied acceleration of the intrinsic carousel rotation cannot easily be explained by plasma models based on \ExB{} drift.
Furthermore, we find that methods of classifying the drift modes by means of $P_3$ measurements can sometimes produce erroneous identifications in the presence of a changing drift rate.
The ``horizontal'' separation between driftbands ($P_2$) is found to be larger at later rotation phases within the pulse window, which is inconsistent with the established effects of retardation, aberration, and the motion of the visible point.
Longer observations spanning at least $\sim 10,000$ pulses are required to determine how the carousel rotation parameters change from one drift sequence to the next.

\end{abstract}

\keywords{pulsars, pulsars --- subpulse drifting, individual pulsars --- \pulsar{}}

\maketitle

%%%%%%%%%%%%%%%%
% INTRODUCTION %
%%%%%%%%%%%%%%%%

\section{Introduction} \label{sec:intro}

Despite nearly half a century of observational and theoretical investigations, the physical mechanisms responsible for the radio emission from pulsars remain unresolved \citep{Michel1991,Beskin1993,Melrose2016}.
Some features of observed pulsar emission are considered vitally important to furthering our understanding of these fundamental processes.
Chief among these is the phenomenon of subpulse drifting, which is the systematic shift in pulse phase of substructures within individual pulses over time \citep{Drake1968}.
Also critical is nulling, in which the radio emission appears to switch off temporarily \citep{Backer1970,Ritchings1976}, a widespread phenomenon closely linked to the coherent emission process.
\pulsar{} is a bright, long-period ($P_1=0.943\,$s) pulsar with dispersion measure $\text{DM}=10.9\,$pc$\,$cm$^{-3}$ that exhibits both subpulse drifting \citep{Backer1970,Huguenin1970} and extensive nulling ($\sim 45\%$ of the time, \citealt{Vivekanand1995}), and thus is an important object which can potentially reveal vital clues to the underlying emission mechanisms.

Subpulses are thought to represent ``subbeams'' caused by discrete emission regions that are stable over many pulsar rotations, which in some cases may be arranged in a ``carousel'' pattern centered on the magnetic dipole axis and rotating around it at some rate, $D$ \citep{Ruderman1975,Rankin1986}.
Strong observational support for this view came with the work of \citet{Deshpande1999,Deshpande2000}, who were able to determine that the regular drifter PSR B0943+10, has a stable carousel consisting of 20 discrete emission regions.
One of the earliest and most successful emission models \citep{Ruderman1975} invoked \ExB{} drift to explain the carousel's circular motion; however, there are some outstanding issues with its quantitative predictions.
First, the measured drift rate for at least some pulsars are known to be many times smaller than that predicted by \ExB{} drift \citep[e.g.][]{Mitra2008}.
Second, the drift rate of some pulsars is not constant, but varies over time.
The most common manifestation of this is the presence of temporally distinct drift modes, characterized by an abrupt change in the drift rate, usually with a timescale less than a single stellar rotation \citep[e.g.][]{Redman2005}.
Variations in drift rate can also occur over longer timescales, without sudden drift modes changes, \citep[e.g.][]{Biggs1985,Bhattacharyya2009}.
\pulsar{} exhibits both long and short timescale drift rate variations, with three distinct drift modes designated as Modes A, B, and C \citep{Huguenin1970,Wright1981}, and with drift rate variations occurring within each mode \citep{Vivekanand1997}.

The methodology of \citet{Deshpande2000} involves mapping the intensity sequences from the drifting subpulses onto a coordinate system centered on the magnetic axis (the so-called ``cartographic transform''),
Ideally, this can be applied to other drifters, such as \pulsar{}, allowing us to determine the geometry and dynamics of the emission regions.
This is not always possible, mainly due to the difficulty in resolving the presence of aliasing, in which the true carousel drift rate is different from the measured drift rate because of the sub-Nyquist sampling of the emitting region due to the star's rotation.
A carousel with an integer number of subbeams admits only a discrete (but possibly infinite) set of solutions to any given observed driftband pattern, with higher drift rates corresponding to higher order aliasing.
Determining the true carousel rate, and hence the order of aliasing present, is difficult because different sets of parameter values (viewing geometry, as well as aliasing) can give rise to identical-looking driftbands.
This is especially true for \pulsar{}, whose multiple drift modes complicate the issue, and whose viewing geometry is not precisely known \citep{Smits2007}.
Nevertheless, it is vital to resolve the aliasing order so we can understand the configuration of emission regions of this pulsar and the relationship between its three different drift modes.

\begin{figure}[!t]
    %\centering
    \input{annotatedpulsestack.tex}
    \caption{A subset of the stacked pulses of \pulsar{}, from MWA observations at $185\,\MHz$. The driftbands are clearly visible. The horizontal and vertical separations between consecutive driftbands (indicated by the overlaid white arrows) are $P_2$, measured in degrees, and $P_3$, measured in units of $P_1$, respectively. Two of the three drift modes, A and B, are exemplified here, distinguished by markedly different $P_3$ values and drift rates. A typical null sequence is also present.}
    \label{fig:annotatedpulsestack}
\end{figure}
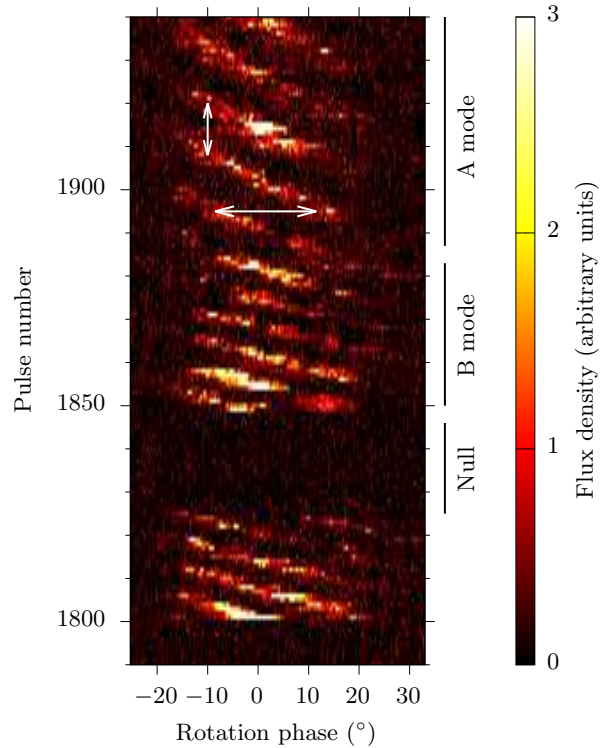

The regularity of both the stellar rotation and of the observed driftbands allows us to define a number of periodicities, often used in subpulse drifting analyses \citep[e.g.][]{Deshpande1999,Edwards2002}.
$P_1$, $P_2$, and $P_3$ are respectively defined as (1) the pulsar's rotation period, (2) the temporal separation between two subpulses within a single rotation, and (3) the time it takes for a subpulse to arrive at the same rotation phase as its predecessor.
The drifting subpulses appear as diagonal \emph{driftbands} in the pulse stack, which is the (one-dimensional) time series plotted in a two-dimensional array, with each row corresponding to $360^\circ$ of rotation, and time progressing along the vertical axis (Fig. \ref{fig:annotatedpulsestack}).
Visually, $P_2$ and $P_3$ are realized as the horizontal and vertical separations (respectively) of consecutive driftbands.
The slope of the driftbands is the observed drift rate, defined as
\begin{equation}
    D \equiv \dv{\phase}{\tvar} = \pm\frac{P_2}{P_3},
    \label{eqn:driftrate}
\end{equation}
where $\phase$ is the phase, $p$ is the pulse number, and the sign indicates the drift direction.

Drift modes are usually characterized by their $P_3$ value, which is generally found to remain stable over the course of a drift sequence (defined here as a set of contiguous pulses all belonging to the same drift mode).
Early work on \pulsar{} measured $P_3$ values of $12.5$, $6.8$, and $4.5\,P_1$ for Modes A, B, and C respectively \citep{Huguenin1970,Wright1981}, with individual drift sequences lasting from a few to hundreds of pulse periods \citep{Vivekanand1997}.
Interestingly, the subpulse phases appear to be correlated across the nulls \citep{Joshi2000}, indicating organized motion of the supposed carousel configuration even during nulls.

\citet{Smits2005,Smits2007} investigated the behavior of these drift modes at nine frequencies ranging from $157\,\MHz$ to $4.85\,\GHz$.
Their analysis revealed that the driftbands of Mode B were largely undetectable at the higher frequency.
They offered a geometric interpretation, placing the surface plasma events during Mode B emission closer to the magnetic pole than the Mode A plasma events.
The assumed radius-to-frequency mapping implies that at higher frequencies, the Mode B emission forms too narrow an emission cone to intersect the line of sight.
The physical cause of the change of magnetic latitude of the surface plasma events, however, was not explored.

Indeed, no physical explanation has been offered for several key properties of \pulsar{}'s drift modes, such as their average duration, the order in which they appear, their relationship to the null sequences, and the variation (albeit small) in their respective $P_3$ measurements.
Crucially, the presence or absence of aliasing in \pulsar{} has not been determined; however, \citet{Smits2007} estimate the number of discrete emissions in the carousel to be about 9, under the assumption that aliasing is not present.

In this paper we present new observations of \pulsar{} made with the Murchison Widefield Array (MWA, \citealt{Tingay2013}), a low frequency precursor to the Square Kilometre Array (SKA).
We attempt to characterize the pulsar's observed drifting behavior in terms of a small number of parameters that remain constant over the course of individual drift sequences.
The data acquisition and pre-processing are described in \S\ref{sec:observations}.
Our analysis of the driftbands, and the investigation of their time-varying behaviour are described in \S\ref{sec:analysis}.
The theoretical implications of a variable drift rate are discussed in \S\ref{sec:discussion}, and our conclusions are presented in \S\ref{sec:conclusion}.

%%%%%%%%%%%%%%%%
% OBSERVATIONS %
%%%%%%%%%%%%%%%%

\section{Observations and data processing} \label{sec:observations}

The data were taken with the MWA, a low frequency aperture array located in remote Western Australia.
The MWA is now geared for high time resolution science, with the recently commissioned Voltage Capture System mode (VCS; \citealt{Tremblay2015}).
The VCS enables the recording of the raw voltages from each of the MWA's 128 tiles, which are downloaded from site to the dedicated data storage facility at the Pawsey Supercomputing Centre\footnote{\url{https://www.pawsey.org.au/}}.

We recorded 42 minutes ($\sim18.5\,\text{TB}$) of VCS data on 19th January 2016.
The data were processed following a procedure similar to \citet{Bhat2016}, which is summarized here.
Calibration of the data was performed with the Real Time System (RTS) software (Mitchell, in prep), using an observation of Pictor A taken immediately prior to the pulsar observation.
Using the calibration solution, the raw voltages were phased up to form a pencil beam ($\sim 2\,$arcmins in diameter) on \pulsar{}.

The resulting data set (stored in the PSRFITS format, \citealt{Hotan2004}) consisted of $24 \times 1.28\,\MHz$ coarse frequency channels ranging from $169.60$ to $200.32\,\MHz$, and a time resolution of $100\,\us$.
The data from only the central $88$ out of $128$ ($10\,\kHz$) fine channels of each coarse channel were kept because of aliasing effects inherent in the polyphase filter bank, which attenuates the response of the antennas at the edges of the coarse channels.
The best solution was found by calibrating on 115 out of the available 128 antenna tiles, so the data from the remaining tiles were rejected from the pulsar analysis.

Finally, the resulting frequency-time data was processed in DSPSR \citep{VanStraten2011b} and PSRCHIVE \citep{Hotan2004} to produce a single-pulse archive and a timeseries.
The time resolution of the timeseries was $0.921\,\ms$, corresponding to $1024$ phase bins across one pulsar period.
The data set is of very high quality, with an average S/N of $\sim 9$ per pulse (without excising null pulses), approximately a factor of $8.5$ times higher than that of the same observation processed incoherently (i.e. with the signal power detected at each tile summed together), $\sim 20\%$ less than the theoretical expectation.
This paper presents the first study of individual pulses for pulsar emission science undertaken with the MWA (but see, e.g., \citealt{Oronsaye2015} for previous single-pulse studies with the MWA).

%%%%%%%%%%%%
% ANALYSIS %
%%%%%%%%%%%%

\section{Subpulse drifting analysis} \label{sec:analysis}

\begin{deluxetable*}{ccccccc}
    \tablecaption{Statistics of drift mode measurements\label{tbl:P3stats}}
    \tablewidth{0pt}
    \tablehead{
        \colhead{Mode} & \colhead{Number of} & \colhead{Mean $P_2$} & \colhead{Mean $P_3$} & \colhead{Mean $P_3$} & \colhead{Occurrence} & \colhead{Mean} \\
        \colhead{} & \colhead{sequences} & \colhead{($^\circ$)} & \colhead{(PAPS) ($P_1$)} & \colhead{(Quad) ($P_1$)} & \colhead{fraction (\%)} & \colhead{duration ($P_1$)}
    }
    \startdata
         A & $  9$ & $ 18.9 \pm   1.1$ & $ 11.9 \pm   2.0$ & $ 12.5 \pm   0.8$ & $ 18.4$ & $ 54.6$ \\
         B & $ 31$ & $ 19.8 \pm   0.5$ & $  7.0 \pm   0.5$ & $  7.0 \pm   0.2$ & $ 34.5$ & $ 29.6$ \\
         C & $  2$ & $ 19.1 \pm   2.9$ & $  5.9 \pm   3.6$ & $  4.6 \pm   0.3$ & $  0.8$ & $ 11.0$ \\
      Null & $ 38$ &  -  &  -  &  -  & $ 45.5$ & $ 31.9$ \\
      Unknown\tablenotemark{a} & $  6$ & $ 19.9 \pm   3.2$ &  -  &  -  & $  0.8$ & $  3.7$ \\
    \enddata
    \tablecomments{The mean $P_3$ values were measured in two ways: (1) the PAPS method of \citet{Smits2005} involves Fourier analysis of pulsestack columns, and is described in \S3.1; (2) the ``Quad'' method involves fitting quadratic lines to sets of driftbands, and is described in \S3.3. The mean errors are Gaussian-propagated from the standard deviations of the same quantities measured for} individual drift sequences.
    \tablenotetext{a}{Sequences that were too short to yield a reliable measurement of $P_3$ were uncategorized.}
\end{deluxetable*}
The analysis in this paper is aimed at exploring how the behaviour of the driftbands of \pulsar{} varies both between different drift modes, and within individual drift sequences.
We analyzed only data within the on-pulse window, which was chosen to fall between the first and last phase bins whose average flux densities were $\sigmathresh\sigma$ above the off-pulse noise (the noise statistics were obtained from phase bins more than $\offpulsebuffer^\circ$ away from the profile peak).
The pulse window was thus determined to be between $\phaselo^\circ \le \phase \le \phasehi^\circ$, where the point $\phase = 0^\circ$ was defined to be the center of the phase bin which contained the largest average flux density.

\subsection{Determining the drift mode boundaries}
\label{sec:initanalysis}

Traditional methods of drift mode analysis include the Harmonic Resolved Fluctuation Spectrum \citep[HRFS;][]{Deshpande2000} and the mathematically equivalent Two-Dimensional Fluctuation Spectrum \citep[2DFS;][]{Edwards2003}.
These methods are designed to measure $P_2$ and $P_3$ by means of Fourier analyses of the time series.
The fluctuation spectra of pulsars with multiple drift modes will contain the Fourier components corresponding to the $(P_2,P_3)$ pairs of each drift sequence and a delocalized component corresponding to the distribution of the drift modes, as can be seen, for example, in the HRFS of PSR B2303+30 \citep{Redman2005}.

Because \pulsar{} exhibits multiple drift modes and long-duration nulls, the components corresponding to the three drift modes are not easily resolved in the fluctuation spectra \citep[e.g., Fig. 8 of ][]{Karuppusamy2011}.
Any realistic measurement of $P_2$ and $P_3$ (and hence the drift rate) can therefore only be achieved by first determining the precise locations of the transitions between different drift modes, and treating each drift sequence separately.
For this pulsar, the drift modes appear to switch on a timescale no longer than a single rotation period, and so it becomes possible to associate each pulse with a distinct mode.

\begin{sidewaysfigure*}[p]
    \centering
    \vspace{9cm}
    \input{P3_DR_P3.tex}
    \caption{\emph{Top}: Distribution of $P_3$ values calculated via the PAPS method of \citet{Smits2005}. \emph{Middle}: The slopes of driftbands from the linear fit described in \S\ref{sec:linearfit}. The horizontal length indicates the extent of the given driftband; the vertical length indicates the error of the drift rate. \emph{Bottom}: The values of continuously changing $P_3$ as predicted by the quadratic fits described in \S\ref{sec:quadfit}. The $P_3$ values of Mode B (green) and Mode C (red) appear to be confined to narrow regions (between $6$ and $8\,P_1$ for B; $3$ and $5\,P_1$ for C), but those of Mode A (blue) vary dramatically. \emph{Color panel}: These are the determined modes, by color: {\color{blue} mode A}, {\color{green} mode B}, {\color{red} mode C}, nulls are in black, unclassified in yellow.}
    \label{fig:P3_DR_P3}
\end{sidewaysfigure*}
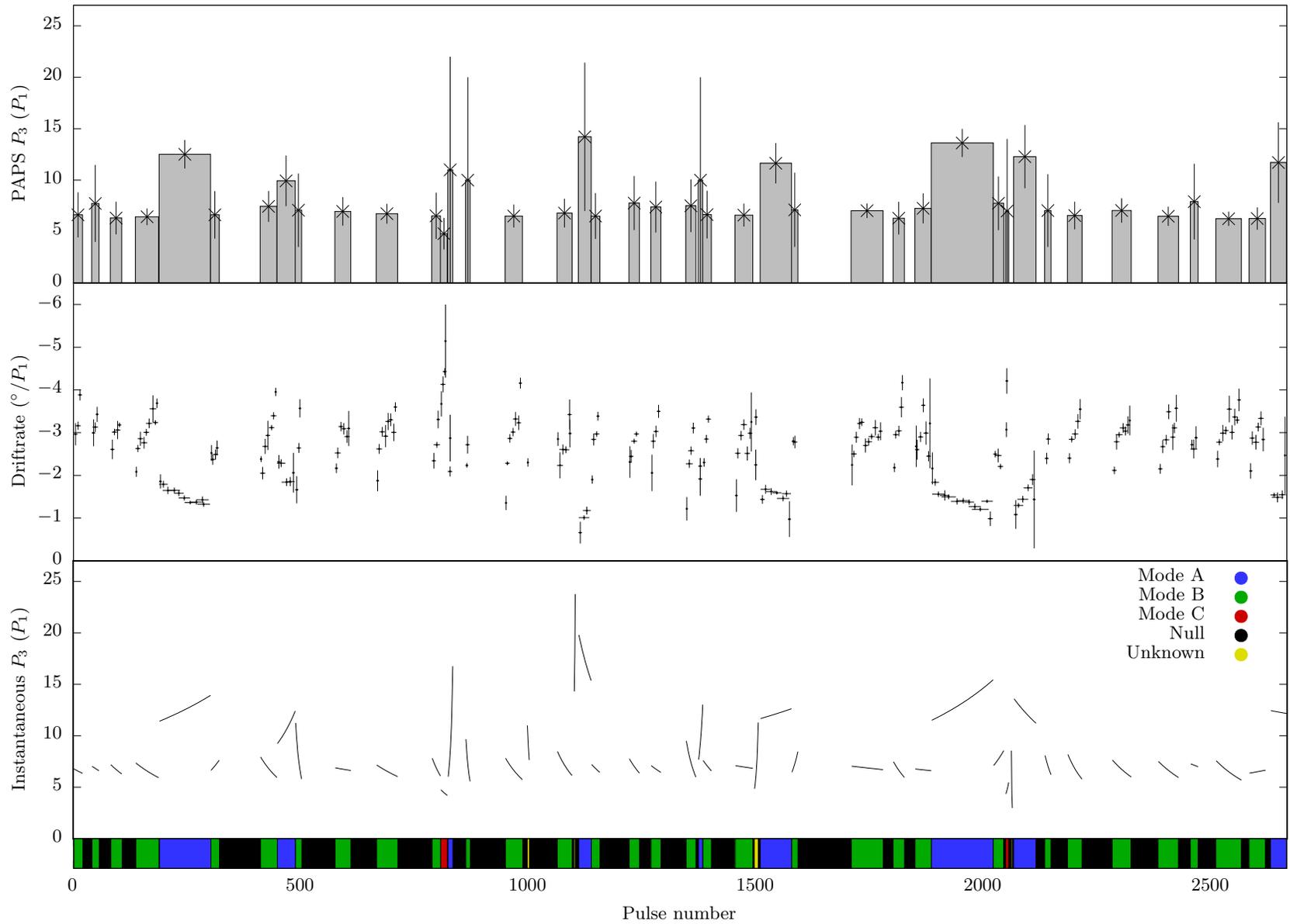

One can use the sliding two-dimensional fluctuation spectrum \citep[S2DFS;][]{Serylak2009} to obtain a map of temporal changes to the drift modes, but the coarse resolution inherent in the technique cannot resolve sudden changes on the time scales of individual pulses.
We therefore followed the method of \citet{Smits2005}, which measures the average $P_3$ in a candidate drift sequence by computing the phase-averaged power spectrum (PAPS), which is the sum of the amplitudes of the DFT of each phase bin.
The beginning and ending boundaries of the drift sequence were then adjusted incrementally until the peak value in the PAPS divided by the rms of the rest of the PAPS was maximized.
The resulting map of drift modes is shown in the top panel of Fig. \ref{fig:P3_DR_P3}, and a summary table of drift mode statistics is given in Table \ref{tbl:P3stats}.

The distribution of drift modes in the MWA observation strongly resembles that of \citet{Smits2005,Smits2007}, who observed the pulsar for a similar length of time at similar frequencies ($157$, $243$, $325\,\MHz$ and above).
In particular, we note the following similarities: Mode A sequences are generally longer than modes B and C, and mode A sequences are often sandwiched between two mode B sequences with minimal nulling between them.

\subsection{Linear fits to driftbands}
\label{sec:linearfit}

The simplest way to characterize individual driftbands is to treat them as independent line segments.
By fitting a line to each driftband, we hope to assess whether there are any systematic changes in the drift rate over the course of a drift sequence.

Each pulse was convolved with a narrow Gaussian ($\text{FWHM} = \convgausswidth^\circ \approx 10\%$ of $P_2$) in order to smooth out high frequency noise fluctuations.
The phase bin containing the most power in each pulse was identified, and the phase of the interpolated peak (using a cubic spline to get a sub-bin estimation) was taken as the phase of a subpulse.
$P_2$ has previously been measured to be $\sim 20^\circ$ at low frequencies (see Fig. \ref{fig:P2historical}), so the second highest peak was identified with the same method, but with the constraint that it was not closer than $\minseparation^\circ$ to the first bin.
For the vast majority of pulses, a maximum of two driftbands were visible in any given pulse, so we did not attempt to find a third subpulse peak.

An algorithm was designed to find connected series of subpulse peaks that belong to the same driftband.
Starting at the beginning of the observation, and assuming that the subpulses found in the first pulse do in fact ``belong'' to genuine driftbands, we assigned the subpulses in the succeeding pulse to the already identified driftbands if the subpulse phases are within $\minseparation^\circ$ of the driftband's projected phase at the pulse in question.
The projected phases were determined by a weighted least squares fit to each subpulse already associated with a driftband, where the weighting was proportional to the peak amplitude of the subpulse.
However, the variation in subpulse position requires that a statistically significant number of pulses be already assigned to a driftband in order to obtain a reliable extrapolation of the driftband location.
Thus, if a driftband has so far only been assigned $\shortseqmaxlength$ subpulses or fewer, we assume a nominal drift rate\footnote{For this pulsar, drift rates appear to range between approximately $-0.5^\circ / P_1$ and $-4.5^\circ / P_1$, as evident in the results.} of $\drestimate^\circ / P_1$ and only perform least squares regression to find the phase offset.
If a subpulse was found with a phase more than $\minseparation^\circ$ to the right of the projected phases, it was assigned to the beginning of a new driftband.
The process continues until the onset of a null sequence, and the entire algorithm is repeated for each drift sequence.
\begin{figure}[!th]
    \centering
    \input{all_lines.tex}
    \caption{Pulse stack showing \pslength{} pulses of \pulsar{}. The image is saturated at $\pssaturationpercent\%$, meaning that any pixel containing a value more than $\pssaturationpercent\%$ of the maximum value in that window is displayed as a black pixel. Overlaid are red lines showing the linear fits to the drift bands using weighted least squares regression, described in \S\ref{sec:linearfit}, and blue lines showing the quadratic curves fit with fewer parameters, described in \S\ref{sec:quadfit}.}
    \label{fig:all_lines}
\end{figure}
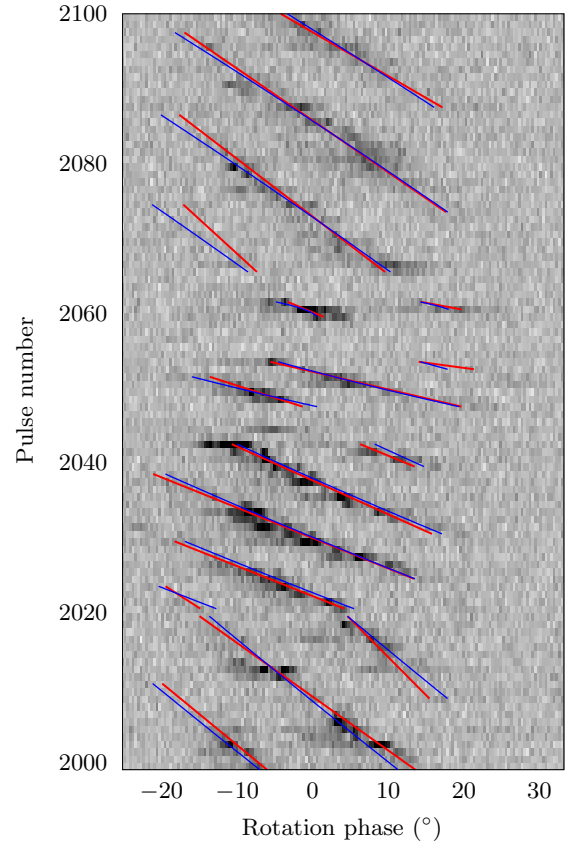
The resulting assignment of subpulses to driftbands was checked by eye for correctness, and the same algorithm was successfully applied to the reverse timeseries to test it for robustness.
Once all determined subpulses had either been assigned to a driftband or rejected as an outlier (if it didn't fall within $\minseparation^\circ$ of any projected driftband), the subpulse positions were fit by a weighted least squares regression as before, except that this was also applied to even short (i.e. containing $\shortseqmaxlength$ or fewer subpulses) driftbands.
The results of this algorithm for a subset of the pulse stack are illustrated by red lines in Fig. \ref{fig:all_lines}.

The errors on the slopes of the driftbands are calculated to be
\begin{equation}
    m_{\text{unc}} =
      \sqrt{\frac{1}{n-2}
            \frac{\sum\limits_i (w_i(\phase_i-\hat{\phase}))^2}{
                  \sum\limits_i (w_i(p_i-\bar{p}))^2}},
\end{equation}
where the sums are iterated over the subpulses within a given driftband; $n$ is the number of pulses within a driftband; $\phase$ and $\hat{\phase}$ are respectively the measured subpulse phase and the phase predicted from the linear fit; $p$ is the pulse number; and $\bar{p} = \frac{1}{n} \sum_i p_i$.

Having obtained a model for each driftband, we can now assess how the driftband slopes vary over the course of individual drift sequences (Fig. \ref{fig:P3_DR_P3}, middle panel).
Mode A driftbands tend to become steeper (i.e. the drift rate decreases), but those of mode B tend to become shallower (i.e. the drift rate increases), but not exclusively.

\subsection{Quadratic fits to driftbands}
\label{sec:quadfit}

The linear fits to the driftbands suggest that the drift rate varies quasi-linearly over the course of each drift sequence (to a first order approximation, cf. middle panel of Fig. \ref{fig:P3_DR_P3}).
Assuming that this is the case, and also assuming that $P_2$ does not vary over time (consistent with what is observed in this data set), we write the following functional form for the $n$th driftband within a given drift sequence:
\begin{equation}
    \phase(\tvar) = a_1\tvar^2 + a_2\tvar + a_3 + a_4n,
    \label{eqn:linparam}
\end{equation}
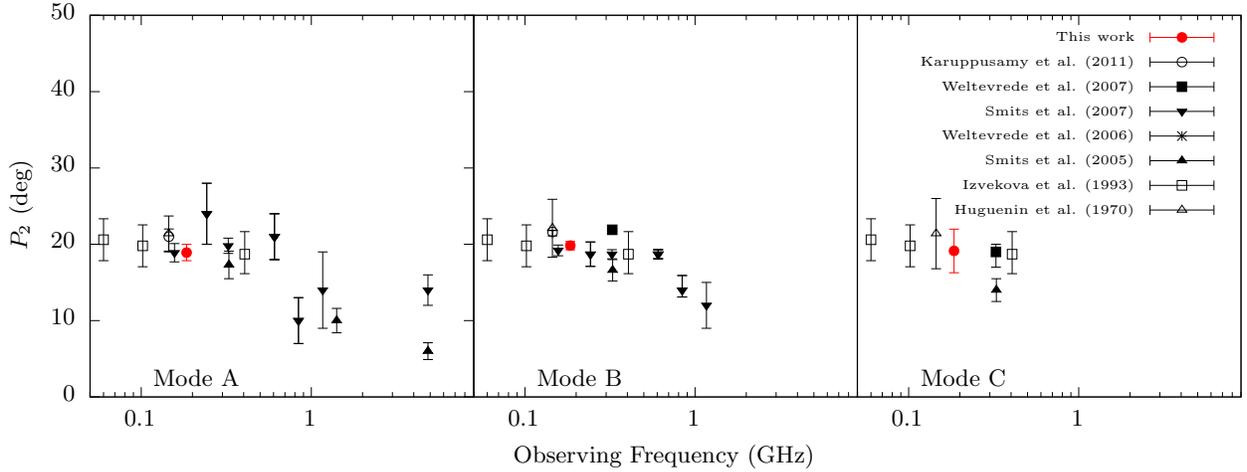
\begin{figure*}[!t]
    \centering
    \input{P2historical.tex}
    \caption{Comparison of available measurements of $P_2$, with the panels from left to right showing measurements of modes A, B, and C respectively. The fact that $P_2$ is consistent with being uniform across all three modes at any given frequency suggests that the angular spacing of the emission regions around the magnetic axis does not change between mode switches.}
    \label{fig:P2historical}
\end{figure*}
\nocite{Karuppusamy2011}
\nocite{Weltevrede2007}
\nocite{Smits2007}
\nocite{Weltevrede2006}
\nocite{Smits2005}
\nocite{Huguenin1970}
where $p$ is the pulse number starting from the beginning of the drift sequence, and $a_1$, $a_2$, $a_3$, and $a_4$ are free parameters to be fit.
The drift rate, $P_2$, and $P_3$ can be derived from these parameters thus:
\begin{equation}
    \begin{gathered}
        \dv{\phase}{\tvar} = 2a_1\tvar + a_2, \\
        P_2 = a_4, \\
        P_3 = \frac{P_2}{\text{d}\phase/\text{d}\tvar}
            = \frac{a_4}{2a_1\tvar + a_2}.
    \end{gathered}
    \label{eqn:quadrelationships}
\end{equation}
Note that the drift rate is linear in $\tvar$, and $P_2$ is constant, as desired.

Because Eq. \eqref{eqn:linparam} is linear in $a_1$, $a_2$, $a_3$, $a_4$, they can be fit to the subpulse position data using least-squares regression.
A weighted\footnote{The weights used were the amplitudes of the subpulse peaks, as before.} linear fit was performed with the subpulses in each drift sequence, resulting in quadratic driftband fits shown in Fig. \ref{fig:all_lines}.

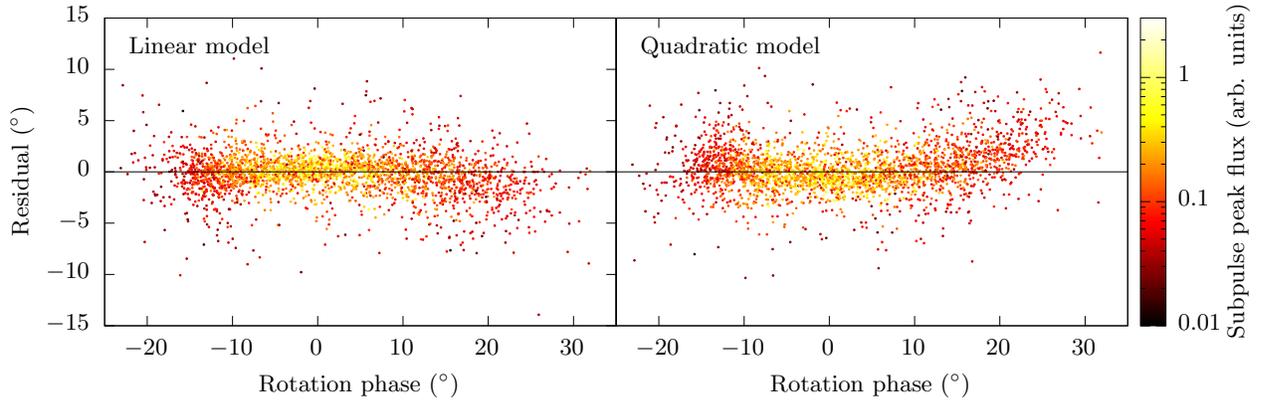
\begin{figure*}[!th]
    \centering
    \input{residual_comparison.tex}
    \caption{The residuals for both the linear fits to the driftbands (left panel) and the quadratic fits (right panel). The color scale indicates the amplitude of the peak of the subpulses, which were used as weights in the regression fits. The linear fits appear to generate slightly smaller residuals, but requires many more free parameters. Both fits perform worse at the edges of the pulse window than in the central region, suggesting that $P_2$ is a weak function of rotation phase.}
    \label{fig:residual_comparison}
\end{figure*}

The quadratic fits described by Eq. \eqref{eqn:linparam} employ four free parameters per drift sequence, while the linear fits employ two free parameters per driftband.
Both types of fit successfully identify the driftbands, but the subpulse position residuals suggest that the linear method fits the subpulses at the extremes of the pulse window slightly better than the quadratic method (Fig. \ref{fig:residual_comparison}).
We note, however, that the quadratic fit method can potentially be used to correct a misidentification of a drift mode and/or drift mode boundaries, which occurred three times in the present data set.
In the first instance, the quadratic fit failed to find a close fit to a set of driftbands when they were erroneously assumed to belong to the same drift mode.
In the second instance, the drift sequence boundaries determined via the PAPS method had to be slightly adjusted in order to produce a good quadratic fit.
Finally, the Mode C drift sequence at approximately pulse number $2050$ was too short for the PAPS method to yield a reliable $P_3$ measurement, but was easily identified as Mode C using the quadratic fit method.
Indeed, the difference between the average $P_3$ values for Mode C shown in Table \ref{tbl:P3stats} and their respective errors is due to the inability of PAPS to deal with such a short sequence.

\subsection{Characterizing $P_2$}

The value of $P_2$ in \pulsar{} is consistent with being constant in time, irrespective of drift mode (\citealt{Smits2005}, but see \citealt{Vivekanand1997} for evidence of the contrary).
However, it has been observed to decrease at higher observing frequencies, in accordance with the radius-to-frequency mapping \citep[see our Fig. \ref{fig:P2historical} and references therein]{Cordes1978}.
Here, we report that $P_2$ is also dependent on the rotation phase, i.e. where the subpulses fall in the pulse window.

\begin{figure*}[!th]
    \centering
    \input{P2s_vs_phase.tex}
    \caption{\emph{Left}: The phase difference of two subpulses ($P_2$ measured for each pulse) against the phase of the midpoint between them. The fact that no $P_2$ was measured below $\minseparation^\circ$ is a consequence of the algorithm used to measure the subpulse phases. \emph{Right}: The value of the $a_4$ parameter ($P_2$ measured for each drift sequence) against the central phase of the swath used in the fit. Both methods indicate that $P_2$ is positively correlated with rotation phase.}
    \label{fig:P2s_vs_phase}
\end{figure*}
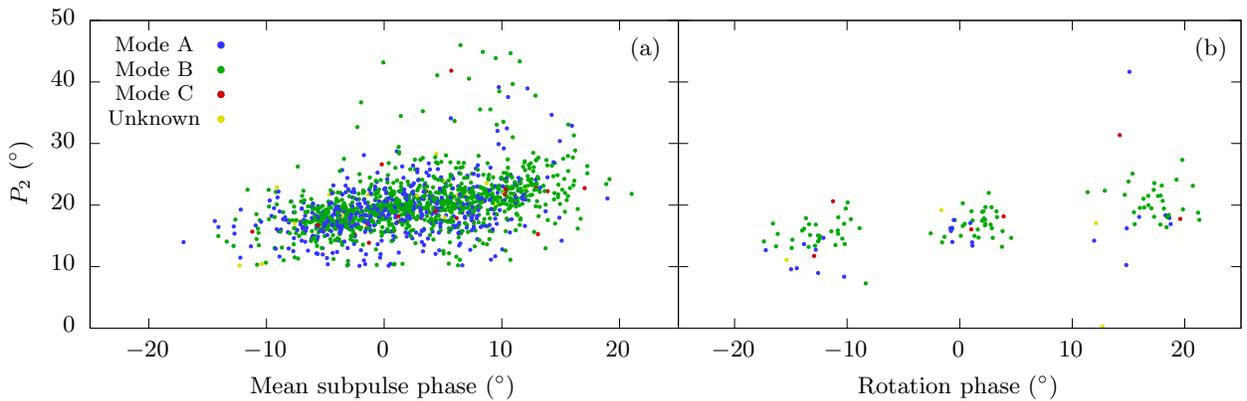

The average $P_2$ value is commonly measured by means of an autocorrelation function applied to the pulse stack.
Here, we measure $P_2$ for each pulse individually, by simply taking the difference of phases of the two subpulse peaks detected by the peak-finding algorithm described in \S\ref{sec:linearfit}.
The resulting $P_2$ measurements are plotted in Fig. \ref{fig:P2s_vs_phase} against the average (absolute) phase of the two subpulses.
There is a noticeable positive correlation between $P_2$ and (average) phase---i.e. subpulses at later phases are generally spaced more widely apart.

To confirm this trend, we performed the identical quadratic fits described above to smaller subsets of subpulses.
Within each drift sequence, the subpulses were divided into three subsets, based on their absolute phases.
The phase boundaries were not the same for each drift sequence; instead, they were chosen to ensure that the number of subpulses in each subset were the same (or differed only by one, if the total number of subpulses was not divisible by three).
$P_2$, as measured by the fit parameter $a_4$ (cf. Eq. \eqref{eqn:linparam}) for each subset is shown plotted against the average phase of the subset in the right-hand panel of Fig. \ref{fig:P2s_vs_phase}.
A similar upward trend is evident.
We note, in passing, that the three modes do not appear to be drawn from different distributions of $P_2$, which is contrary to the finding of \citet{Vivekanand1997} who reported a weakly negative correlation between $P_2$ and drift rate from analysis of their observations at $325\,\MHz$.

%%%%%%%%%%%%%%
% DISCUSSION %
%%%%%%%%%%%%%%

\section{Discussion}
\label{sec:discussion}

We have demonstrated the presence of two effects in \pulsar{} that have previously not been studied in detail.
Firstly, the drift rate in \pulsar{} varies gradually within individual drift modes as well as sharply between them (Fig. \ref{fig:P3_DR_P3}).
Secondly, $P_2$ appears to be positively correlated with rotation phase; i.e. subpulses on the right hand side of the pulse window (when viewed in the pulse stack) are more widely separated than those on the left hand side (Fig. \ref{fig:P2s_vs_phase}).

Both of these effects were observed in \pulsartwo{} \citep[hereafter \GG{}]{Gupta2004}, another long-period ($P_1 = 1.85\,$s) drifter, although different from \pulsar{} in many respects.
\pulsartwo{} has a very wide profile, with up to $13$ distinct driftbands being observed across all $360^\circ$.
It appears to exhibit nulling, but weak emission (at the $2\%$ level, at $1374\,\MHz$) has been detected during null periods, enabling the tracing of the driftbands continuously over several hundred pulses \citep{Esamdin2005}.
\citet{Bhattacharyya2008}, however, report no emission during nulls (at the $1\%$ level) in their observations at $157$, $325$, $610$, and $1060\,\MHz$.
\pulsartwo{} and \pulsar{} are suspected to have similar viewing geometries, with $\alpha$ (the angle between the rotation axis and the magnetic axis) and $\beta$ (the angle between then magnetic axis and the line of sight at closet approach) both being very small ($<10^\circ$).
The parameters $\alpha$ and $\beta$ are often estimated by fitting the polarization angle sweeps to the magnetic pole model \citep{Radhakrishnan1969}.
However, for both pulsars, there are several combinations of $\alpha$ and $\beta$ that reproduced the observed polarization curve, and current estimates give $1^\circ.5 \le \alpha \le 5^\circ.0$ and $0^\circ.6 \le \beta \le 2^\circ.0$ for \pulsartwo{} (\GG{}) and $\alpha \approx \beta$ around $0^\circ.1$ to $6^\circ.0$ \citep{Smits2007}.
A final difference is that, unlike \pulsar{} whose drift rate always maintains the same direction and changes only gradually throughout any given drift sequence, the drift rate of PSR B0826-34 fluctuates about a mean value of $\sim 0^\circ / P_1$, changing sign in a quasi-periodic manner.

\subsection{Variable drift rate by stellar surface temperature fluctuations}

Given the apparent similarity with \pulsartwo{}, we assess if the explanations that \GG{} offer for the appearance of these features in \pulsartwo{} would be applicable to \pulsar{} as well.
The variability of the drift rate is suggested to arise from fluctuations in the stellar surface temperature, which indirectly influences the \ExB{} drift rate.
Moreover, the apparent change in drift direction is attributed to an aliasing effect; with higher-order aliasing, the direction will appear to change if the true value of $P_3$ fluctuates over a small range that straddles an integral multiple of $P_1$.
In this scenario, small fractional variations in the true drift rate can appear to observers as large fractional variations in measured drift rate.
\GG{} calculate that only up to $8\%$ change in the true drift rate is required to explain the measured variation, and that this upper limit requires only a $0.14\%$ fluctuation in the surface temperature.

Even though the observed drift rate in \pulsar{} is ubiquitously unidirectional, the fractional variation is very large ($\approx 67\%$) and can only be realistically explained by the same mechanism of fluctuating surface temperature if higher order aliasing is invoked to bring the true drift rate variation down to a few percent.
For the geometry assumed by \citet{Smits2005}, with 9 subbeams viewed at $\alpha \approx \beta \approx 4.5^\circ$, we find that the true drift rate becomes approximately $(10/k)\%$ for aliasing order $k \ge 1$.
Thus, even aliasing order $k \gtrsim 2$ would bring the true drift rate to a level consistent with the $8\%$ inferred for \pulsartwo{}.
However, this cannot explain both the gradual change in drift rate within drift sequences and the abrupt change in drift rate between them, unless the temperature fluctuated at two distinct timescales.
Even if aliasing is present, one is still left with a drift rate that is discontinuous at mode boundaries (cf. \citealt{Esamdin2005}, who point out the existence of cusps in the drifting pattern of B0826-34).
Thus, in keeping with the observed timescale of drift rate variation observed in B0826-34 ($\sim 100$ pulses $\approx 180\,$s), we surmise that if the surface temperature is indeed responsible for the intra-sequence drift rate variation, then some other, possibly unrelated mechanism must drive the drift mode changes.

It is of interest to know what the relationship is, if any, between the drift rates and carousel acceleration parameters of neighboring drift sequences.
The presence of aliasing would obscure such a relationship, which suggests the possibility that the correct aliasing order and number of subbeams may be recognized by their ability to reveal a natural progression of carousel rotation parameters from one sequence to the next.
This, however, requires a much longer data set than presented here since most drift sequences in \pulsar{} are bordered by nulls (see Fig. \ref{fig:P3_DR_P3}), and it is as yet unclear how the subpulse phase, drift rate, and carousel acceleration evolve during a null \citep{Joshi2000}.
Based on the number of mode transitions in the present data set, we estimate a rate of approximately $40$ suitable mode transitions (i.e. without intervening nulls) in $10,000$ pulses, a conservative minimum required to investigate these relationships further.

\subsection{The $P_2$ dependence on rotation phase}

\GG{} invoke a new idea to explain the positive correlation between the measured $P_2$ and rotation phases.
They suggest that the carousel pattern is centered not on the magnetic dipole axis, but around some other nearby axis (dubbed the ``local pole'') that arises perhaps due to a more dominant multipolar component near the surface.
In their case, they were able to determine that a local pole which is offset from the dipole axis by $\sim 3^\circ$ is able to reproduce the observed $P_2$ dependence on rotation phase.
We hope that a similar analysis will be able to find a local pole solution for \pulsar{}.
This analysis, however, also depends on a known (or assumed) number of beams and aliasing order, which cannot yet be inferred from current observations.

Other explanations for this effect are not forthcoming.
For example, aberration effects are traditionally invoked to explain the asymmetry in pulsar profiles in which conal emission features are shifted to lower phases relative to core emission.
Presumably, such effects are present in PSR J0034-0721, but due to the fact that its profile contains only a single (conal) component, they would not be detectable in the traditional way.
However, any measurable feature with a dependence on rotation phase (such as $P_2$), should be ``stretched'' out at lower phases and ``compressed'' at higher phases, similarly to profile components \citep{Gupta2003,Dyks2003}.
However, exactly the opposite is observed in the $P_2$ of \pulsar{}, so aberration effects cannot be invoked to explain it.

Another possible explanation is the motion of the visible point as discussed by \citet{Yuen2014} and \citet{Yuen2016}, which takes into account the direction of the dipolar magnetic field at the emission site, an effect that has traditionally been neglected in the interpretation of $P_2$ measurements.
They showed that measurements of $P_2$ can dramatically underestimate the true subbeam separation when the angle between the rotation and magnetic axes, $\alpha$, is sufficiently small, which is believed to be the case for \pulsar{} \citep[e.g.][]{Smits2005,Smits2007}.
However, they also showed that the discrepancy in $P_2$ is symmetrical about the fiducial point, with the measured $P_2$ increasing as one moves away (in either direction) from the fiducial point.
Again, this is at odds with what is seen from our observations, where $P_2$ appears to increase monotonically across the pulse window.
Even though invoking the effect of a moving visible point may explain the variation in $P_2$, it will require the assumption that the fiducial point lies somewhere to the left of the on-pulse region.
However, attempts to constrain the fiducial point of \pulsar{} have not met with success because of the difficulty of locating a zero-crossing point in the position-angle sweep of the dominant polarization, due to its flatness over the pulse window \citep{Smits2007}.
Thus we are not able to comment on the likelihood of this scenario.

%%%%%%%%%%%%%%
% CONCLUSION %
%%%%%%%%%%%%%%

\section{Conclusion} \label{sec:conclusion}

We have conducted a detailed analysis of new observations of \pulsar{} with the MWA at $185\,\MHz$, a first-of-its-kind demonstration of this instrument's capability of producing high quality single-pulse data, in line with its intended science aims \citep{Bowman2013}.
Our analysis shows that the driftbands of \pulsar{} exhibit more complex behavior than what has been inferred from previous studies.
In particular, (1) the measured drift rate changes continuously within individual drift sequences, with a characteristic variation time scale apparently longer than the typical duration of individual drift sequences; and (2) $P_2$ is positively correlated with rotation phase.
Both of these effects were observed and studied in \pulsartwo{} by \citet{Gupta2004}, who explain the variable drift rate by linking it to surface temperature fluctuations, and the $P_2$ dependence on rotation phase by determining the position of a ``local pole'' around which the carousel is assumed to rotate.
However, the applicability of their proposed physical explanations to \pulsar{} requires the knowledge of the aliasing order and the number of subbeams in the carousel, which are currently unknown for \pulsar{}.
However, we note that resolving the aliasing order and number of subbeams may be helped by assuming that the true drift rate varies continuously over drift mode boundaries, but such an investigation requires significantly longer observations than have been presented here.

%%%%%%%%%%%%%%
% END MATTER %
%%%%%%%%%%%%%%

\subsubsection*{Acknowledgements}

This scientific work makes use of the Murchison Radio-astronomy Observatory, operated by CSIRO.
We acknowledge the Wajarri Yamatji people as the traditional owners of the Observatory site.
Support for the operation of the MWA is provided by the Australian Government (NCRIS), under a contract to Curtin University administered by Astronomy Australia Limited.
We acknowledge the Pawsey Supercomputing Centre which is supported by the Western Australian and Australian Governments.
We would also like to thank the referee, Patrick Weltevrede, for the detailed comments and suggestions which improved the paper.

Parts of this research were conducted by the Australian Research Council Centre of Excellence for All-sky Astrophysics (CAASTRO), through project number CE110001020.
NDRB acknowledges the support from a Curtin Research Fellowship (CRF12228).

\facility{MWA}

\bibliography{biblio}

\end{document}

%% file: annotatedpulsestack.tex
% GNUPLOT: LaTeX picture with Postscript
\begingroup
  \makeatletter
  \providecommand\color[2][]{%
    \GenericError{(gnuplot) \space\space\space\@spaces}{%
      Package color not loaded in conjunction with
      terminal option `colourtext'%
    }{See the gnuplot documentation for explanation.%
    }{Either use 'blacktext' in gnuplot or load the package
      color.sty in LaTeX.}%
    \renewcommand\color[2][]{}%
  }%
  \providecommand\includegraphics[2][]{%
    \GenericError{(gnuplot) \space\space\space\@spaces}{%
      Package graphicx or graphics not loaded%
    }{See the gnuplot documentation for explanation.%
    }{The gnuplot epslatex terminal needs graphicx.sty or graphics.sty.}%
    \renewcommand\includegraphics[2][]{}%
  }%
  \providecommand\rotatebox[2]{#2}%
  \@ifundefined{ifGPcolor}{%
    \newif\ifGPcolor
    \GPcolortrue
  }{}%
  \@ifundefined{ifGPblacktext}{%
    \newif\ifGPblacktext
    \GPblacktexttrue
  }{}%
  % define a \g@addto@macro without @ in the name:
  \let\gplgaddtomacro\g@addto@macro
  % define empty templates for all commands taking text:
  \gdef\gplbacktext{}%
  \gdef\gplfronttext{}%
  \makeatother
  \ifGPblacktext
    % no textcolor at all
    \def\colorrgb#1{}%
    \def\colorgray#1{}%
  \else
    % gray or color?
    \ifGPcolor
      \def\colorrgb#1{\color[rgb]{#1}}%
      \def\colorgray#1{\color[gray]{#1}}%
      \expandafter\def\csname LTw\endcsname{\color{white}}%
      \expandafter\def\csname LTb\endcsname{\color{black}}%
      \expandafter\def\csname LTa\endcsname{\color{black}}%
      \expandafter\def\csname LT0\endcsname{\color[rgb]{1,0,0}}%
      \expandafter\def\csname LT1\endcsname{\color[rgb]{0,1,0}}%
      \expandafter\def\csname LT2\endcsname{\color[rgb]{0,0,1}}%
      \expandafter\def\csname LT3\endcsname{\color[rgb]{1,0,1}}%
      \expandafter\def\csname LT4\endcsname{\color[rgb]{0,1,1}}%
      \expandafter\def\csname LT5\endcsname{\color[rgb]{1,1,0}}%
      \expandafter\def\csname LT6\endcsname{\color[rgb]{0,0,0}}%
      \expandafter\def\csname LT7\endcsname{\color[rgb]{1,0.3,0}}%
      \expandafter\def\csname LT8\endcsname{\color[rgb]{0.5,0.5,0.5}}%
    \else
      % gray
      \def\colorrgb#1{\color{black}}%
      \def\colorgray#1{\color[gray]{#1}}%
      \expandafter\def\csname LTw\endcsname{\color{white}}%
      \expandafter\def\csname LTb\endcsname{\color{black}}%
      \expandafter\def\csname LTa\endcsname{\color{black}}%
      \expandafter\def\csname LT0\endcsname{\color{black}}%
      \expandafter\def\csname LT1\endcsname{\color{black}}%
      \expandafter\def\csname LT2\endcsname{\color{black}}%
      \expandafter\def\csname LT3\endcsname{\color{black}}%
      \expandafter\def\csname LT4\endcsname{\color{black}}%
      \expandafter\def\csname LT5\endcsname{\color{black}}%
      \expandafter\def\csname LT6\endcsname{\color{black}}%
      \expandafter\def\csname LT7\endcsname{\color{black}}%
      \expandafter\def\csname LT8\endcsname{\color{black}}%
    \fi
  \fi
    \setlength{\unitlength}{0.0500bp}%
    \ifx\gptboxheight\undefined%
      \newlength{\gptboxheight}%
      \newlength{\gptboxwidth}%
      \newsavebox{\gptboxtext}%
    \fi%
    \setlength{\fboxrule}{0.5pt}%
    \setlength{\fboxsep}{1pt}%
\begin{picture}(4063.00,5760.00)%
    \gplgaddtomacro\gplbacktext{%
      \csname LTb\endcsname%
      \put(747,995){\makebox(0,0)[r]{\strut{}$1800$}}%
      \csname LTb\endcsname%
      \put(747,2623){\makebox(0,0)[r]{\strut{}$1850$}}%
      \csname LTb\endcsname%
      \put(747,4252){\makebox(0,0)[r]{\strut{}$1900$}}%
      \csname LTb\endcsname%
      \put(1125,409){\makebox(0,0){\strut{}$-20$}}%
      \csname LTb\endcsname%
      \put(1506,409){\makebox(0,0){\strut{}$-10$}}%
      \csname LTb\endcsname%
      \put(1886,409){\makebox(0,0){\strut{}$0$}}%
      \csname LTb\endcsname%
      \put(2267,409){\makebox(0,0){\strut{}$10$}}%
      \csname LTb\endcsname%
      \put(2647,409){\makebox(0,0){\strut{}$20$}}%
      \csname LTb\endcsname%
      \put(3028,409){\makebox(0,0){\strut{}$30$}}%
      \csname LTb\endcsname%
      \put(3485,4643){\rotatebox{90}{\makebox(0,0){\strut{}A mode}}}%
      \csname LTb\endcsname%
      \put(3485,3145){\rotatebox{90}{\makebox(0,0){\strut{}B mode}}}%
      \csname LTb\endcsname%
      \put(3485,2135){\rotatebox{90}{\makebox(0,0){\strut{}Null}}}%
    }%
    \gplgaddtomacro\gplfronttext{%
      \csname LTb\endcsname%
      \put(144,3112){\rotatebox{-270}{\makebox(0,0){\strut{}Pulse number}}}%
      \csname LTb\endcsname%
      \put(2033,130){\makebox(0,0){\strut{}Rotation phase ($^\circ$)}}%
      \csname LTb\endcsname%
      \put(4074,669){\makebox(0,0)[l]{\strut{}$0$}}%
      \csname LTb\endcsname%
      \put(4074,2297){\makebox(0,0)[l]{\strut{}$1$}}%
      \csname LTb\endcsname%
      \put(4074,3926){\makebox(0,0)[l]{\strut{}$2$}}%
      \csname LTb\endcsname%
      \put(4074,5555){\makebox(0,0)[l]{\strut{}$3$}}%
      \csname LTb\endcsname%
      \put(4380,3112){\rotatebox{-270}{\makebox(0,0){\strut{}Flux density (arbitrary units)}}}%
    }%
    \gplbacktext
    \put(0,0){\includegraphics{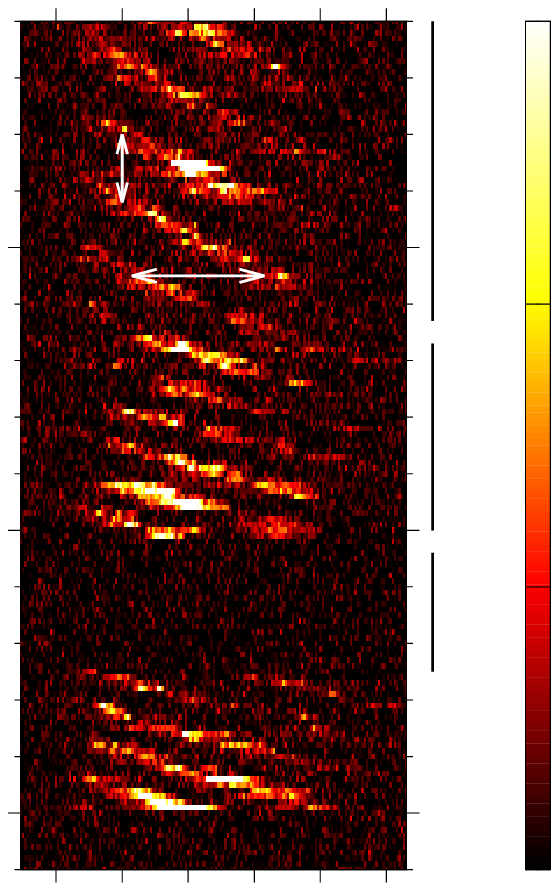}}%
    \gplfronttext
  \end{picture}%
\endgroup

%% file: P3_DR_P3.tex
% GNUPLOT: LaTeX picture with Postscript
\begingroup
  \makeatletter
  \providecommand\color[2][]{%
    \GenericError{(gnuplot) \space\space\space\@spaces}{%
      Package color not loaded in conjunction with
      terminal option `colourtext'%
    }{See the gnuplot documentation for explanation.%
    }{Either use 'blacktext' in gnuplot or load the package
      color.sty in LaTeX.}%
    \renewcommand\color[2][]{}%
  }%
  \providecommand\includegraphics[2][]{%
    \GenericError{(gnuplot) \space\space\space\@spaces}{%
      Package graphicx or graphics not loaded%
    }{See the gnuplot documentation for explanation.%
    }{The gnuplot epslatex terminal needs graphicx.sty or graphics.sty.}%
    \renewcommand\includegraphics[2][]{}%
  }%
  \providecommand\rotatebox[2]{#2}%
  \@ifundefined{ifGPcolor}{%
    \newif\ifGPcolor
    \GPcolortrue
  }{}%
  \@ifundefined{ifGPblacktext}{%
    \newif\ifGPblacktext
    \GPblacktexttrue
  }{}%
  % define a \g@addto@macro without @ in the name:
  \let\gplgaddtomacro\g@addto@macro
  % define empty templates for all commands taking text:
  \gdef\gplbacktext{}%
  \gdef\gplfronttext{}%
  \makeatother
  \ifGPblacktext
    % no textcolor at all
    \def\colorrgb#1{}%
    \def\colorgray#1{}%
  \else
    % gray or color?
    \ifGPcolor
      \def\colorrgb#1{\color[rgb]{#1}}%
      \def\colorgray#1{\color[gray]{#1}}%
      \expandafter\def\csname LTw\endcsname{\color{white}}%
      \expandafter\def\csname LTb\endcsname{\color{black}}%
      \expandafter\def\csname LTa\endcsname{\color{black}}%
      \expandafter\def\csname LT0\endcsname{\color[rgb]{1,0,0}}%
      \expandafter\def\csname LT1\endcsname{\color[rgb]{0,1,0}}%
      \expandafter\def\csname LT2\endcsname{\color[rgb]{0,0,1}}%
      \expandafter\def\csname LT3\endcsname{\color[rgb]{1,0,1}}%
      \expandafter\def\csname LT4\endcsname{\color[rgb]{0,1,1}}%
      \expandafter\def\csname LT5\endcsname{\color[rgb]{1,1,0}}%
      \expandafter\def\csname LT6\endcsname{\color[rgb]{0,0,0}}%
      \expandafter\def\csname LT7\endcsname{\color[rgb]{1,0.3,0}}%
      \expandafter\def\csname LT8\endcsname{\color[rgb]{0.5,0.5,0.5}}%
    \else
      % gray
      \def\colorrgb#1{\color{black}}%
      \def\colorgray#1{\color[gray]{#1}}%
      \expandafter\def\csname LTw\endcsname{\color{white}}%
      \expandafter\def\csname LTb\endcsname{\color{black}}%
      \expandafter\def\csname LTa\endcsname{\color{black}}%
      \expandafter\def\csname LT0\endcsname{\color{black}}%
      \expandafter\def\csname LT1\endcsname{\color{black}}%
      \expandafter\def\csname LT2\endcsname{\color{black}}%
      \expandafter\def\csname LT3\endcsname{\color{black}}%
      \expandafter\def\csname LT4\endcsname{\color{black}}%
      \expandafter\def\csname LT5\endcsname{\color{black}}%
      \expandafter\def\csname LT6\endcsname{\color{black}}%
      \expandafter\def\csname LT7\endcsname{\color{black}}%
      \expandafter\def\csname LT8\endcsname{\color{black}}%
    \fi
  \fi
    \setlength{\unitlength}{0.0500bp}%
    \ifx\gptboxheight\undefined%
      \newlength{\gptboxheight}%
      \newlength{\gptboxwidth}%
      \newsavebox{\gptboxtext}%
    \fi%
    \setlength{\fboxrule}{0.5pt}%
    \setlength{\fboxsep}{1pt}%
\begin{picture}(12960.00,9640.00)%
    \gplgaddtomacro\gplbacktext{%
      \csname LTb\endcsname%
      \put(805,6651){\makebox(0,0)[r]{\strut{}$0$}}%
      \csname LTb\endcsname%
      \put(805,7151){\makebox(0,0)[r]{\strut{}$5$}}%
      \csname LTb\endcsname%
      \put(805,7650){\makebox(0,0)[r]{\strut{}$10$}}%
      \csname LTb\endcsname%
      \put(805,8150){\makebox(0,0)[r]{\strut{}$15$}}%
      \csname LTb\endcsname%
      \put(805,8650){\makebox(0,0)[r]{\strut{}$20$}}%
      \csname LTb\endcsname%
      \put(805,9149){\makebox(0,0)[r]{\strut{}$25$}}%
    }%
    \gplgaddtomacro\gplfronttext{%
      \csname LTb\endcsname%
      \put(406,8000){\rotatebox{-270}{\makebox(0,0){\strut{}PAPS $P_3$ ($P_1$)}}}%
    }%
    \gplgaddtomacro\gplbacktext{%
      \csname LTb\endcsname%
      \put(805,6442){\makebox(0,0)[r]{\strut{}$-6$}}%
      \csname LTb\endcsname%
      \put(805,6027){\makebox(0,0)[r]{\strut{}$-5$}}%
      \csname LTb\endcsname%
      \put(805,5612){\makebox(0,0)[r]{\strut{}$-4$}}%
      \csname LTb\endcsname%
      \put(805,5197){\makebox(0,0)[r]{\strut{}$-3$}}%
      \csname LTb\endcsname%
      \put(805,4782){\makebox(0,0)[r]{\strut{}$-2$}}%
      \csname LTb\endcsname%
      \put(805,4367){\makebox(0,0)[r]{\strut{}$-1$}}%
      \csname LTb\endcsname%
      \put(805,3952){\makebox(0,0)[r]{\strut{}$0$}}%
    }%
    \gplgaddtomacro\gplfronttext{%
      \csname LTb\endcsname%
      \put(406,5301){\rotatebox{-270}{\makebox(0,0){\strut{}Driftrate ($^\circ$/$P_1$)}}}%
    }%
    \gplgaddtomacro\gplbacktext{%
      \csname LTb\endcsname%
      \put(805,1253){\makebox(0,0)[r]{\strut{}$0$}}%
      \csname LTb\endcsname%
      \put(805,1753){\makebox(0,0)[r]{\strut{}$5$}}%
      \csname LTb\endcsname%
      \put(805,2252){\makebox(0,0)[r]{\strut{}$10$}}%
      \csname LTb\endcsname%
      \put(805,2752){\makebox(0,0)[r]{\strut{}$15$}}%
      \csname LTb\endcsname%
      \put(805,3252){\makebox(0,0)[r]{\strut{}$20$}}%
      \csname LTb\endcsname%
      \put(805,3751){\makebox(0,0)[r]{\strut{}$25$}}%
    }%
    \gplgaddtomacro\gplfronttext{%
      \csname LTb\endcsname%
      \put(406,2602){\rotatebox{-270}{\makebox(0,0){\strut{}Instantaneous $P_3$ ($P_1$)}}}%
    }%
    \gplgaddtomacro\gplbacktext{%
    }%
    \gplgaddtomacro\gplfronttext{%
      \csname LTb\endcsname%
      \put(11911,3784){\makebox(0,0)[r]{\strut{}Mode A}}%
      \csname LTb\endcsname%
      \put(11911,3598){\makebox(0,0)[r]{\strut{}Mode B}}%
      \csname LTb\endcsname%
      \put(11911,3412){\makebox(0,0)[r]{\strut{}Mode C}}%
      \csname LTb\endcsname%
      \put(11911,3226){\makebox(0,0)[r]{\strut{}Null}}%
      \csname LTb\endcsname%
      \put(11911,3040){\makebox(0,0)[r]{\strut{}Unknown}}%
    }%
    \gplgaddtomacro\gplbacktext{%
    }%
    \gplgaddtomacro\gplfronttext{%
    }%
    \gplgaddtomacro\gplbacktext{%
    }%
    \gplgaddtomacro\gplfronttext{%
    }%
    \gplgaddtomacro\gplbacktext{%
    }%
    \gplgaddtomacro\gplfronttext{%
    }%
    \gplgaddtomacro\gplbacktext{%
    }%
    \gplgaddtomacro\gplfronttext{%
    }%
    \gplgaddtomacro\gplbacktext{%
    }%
    \gplgaddtomacro\gplfronttext{%
    }%
    \gplgaddtomacro\gplbacktext{%
    }%
    \gplgaddtomacro\gplfronttext{%
    }%
    \gplgaddtomacro\gplbacktext{%
    }%
    \gplgaddtomacro\gplfronttext{%
    }%
    \gplgaddtomacro\gplbacktext{%
    }%
    \gplgaddtomacro\gplfronttext{%
    }%
    \gplgaddtomacro\gplbacktext{%
    }%
    \gplgaddtomacro\gplfronttext{%
    }%
    \gplgaddtomacro\gplbacktext{%
    }%
    \gplgaddtomacro\gplfronttext{%
    }%
    \gplgaddtomacro\gplbacktext{%
    }%
    \gplgaddtomacro\gplfronttext{%
    }%
    \gplgaddtomacro\gplbacktext{%
    }%
    \gplgaddtomacro\gplfronttext{%
    }%
    \gplgaddtomacro\gplbacktext{%
    }%
    \gplgaddtomacro\gplfronttext{%
    }%
    \gplgaddtomacro\gplbacktext{%
    }%
    \gplgaddtomacro\gplfronttext{%
    }%
    \gplgaddtomacro\gplbacktext{%
    }%
    \gplgaddtomacro\gplfronttext{%
    }%
    \gplgaddtomacro\gplbacktext{%
    }%
    \gplgaddtomacro\gplfronttext{%
    }%
    \gplgaddtomacro\gplbacktext{%
    }%
    \gplgaddtomacro\gplfronttext{%
    }%
    \gplgaddtomacro\gplbacktext{%
    }%
    \gplgaddtomacro\gplfronttext{%
    }%
    \gplgaddtomacro\gplbacktext{%
    }%
    \gplgaddtomacro\gplfronttext{%
    }%
    \gplgaddtomacro\gplbacktext{%
    }%
    \gplgaddtomacro\gplfronttext{%
    }%
    \gplgaddtomacro\gplbacktext{%
    }%
    \gplgaddtomacro\gplfronttext{%
    }%
    \gplgaddtomacro\gplbacktext{%
    }%
    \gplgaddtomacro\gplfronttext{%
    }%
    \gplgaddtomacro\gplbacktext{%
    }%
    \gplgaddtomacro\gplfronttext{%
    }%
    \gplgaddtomacro\gplbacktext{%
    }%
    \gplgaddtomacro\gplfronttext{%
    }%
    \gplgaddtomacro\gplbacktext{%
    }%
    \gplgaddtomacro\gplfronttext{%
    }%
    \gplgaddtomacro\gplbacktext{%
    }%
    \gplgaddtomacro\gplfronttext{%
    }%
    \gplgaddtomacro\gplbacktext{%
    }%
    \gplgaddtomacro\gplfronttext{%
    }%
    \gplgaddtomacro\gplbacktext{%
    }%
    \gplgaddtomacro\gplfronttext{%
    }%
    \gplgaddtomacro\gplbacktext{%
    }%
    \gplgaddtomacro\gplfronttext{%
    }%
    \gplgaddtomacro\gplbacktext{%
    }%
    \gplgaddtomacro\gplfronttext{%
    }%
    \gplgaddtomacro\gplbacktext{%
    }%
    \gplgaddtomacro\gplfronttext{%
    }%
    \gplgaddtomacro\gplbacktext{%
    }%
    \gplgaddtomacro\gplfronttext{%
    }%
    \gplgaddtomacro\gplbacktext{%
    }%
    \gplgaddtomacro\gplfronttext{%
    }%
    \gplgaddtomacro\gplbacktext{%
    }%
    \gplgaddtomacro\gplfronttext{%
    }%
    \gplgaddtomacro\gplbacktext{%
    }%
    \gplgaddtomacro\gplfronttext{%
    }%
    \gplgaddtomacro\gplbacktext{%
    }%
    \gplgaddtomacro\gplfronttext{%
    }%
    \gplgaddtomacro\gplbacktext{%
    }%
    \gplgaddtomacro\gplfronttext{%
    }%
    \gplgaddtomacro\gplbacktext{%
    }%
    \gplgaddtomacro\gplfronttext{%
    }%
    \gplgaddtomacro\gplbacktext{%
    }%
    \gplgaddtomacro\gplfronttext{%
    }%
    \gplgaddtomacro\gplbacktext{%
    }%
    \gplgaddtomacro\gplfronttext{%
    }%
    \gplgaddtomacro\gplbacktext{%
    }%
    \gplgaddtomacro\gplfronttext{%
    }%
    \gplgaddtomacro\gplbacktext{%
    }%
    \gplgaddtomacro\gplfronttext{%
    }%
    \gplgaddtomacro\gplbacktext{%
    }%
    \gplgaddtomacro\gplfronttext{%
    }%
    \gplgaddtomacro\gplbacktext{%
    }%
    \gplgaddtomacro\gplfronttext{%
    }%
    \gplgaddtomacro\gplbacktext{%
    }%
    \gplgaddtomacro\gplfronttext{%
    }%
    \gplgaddtomacro\gplbacktext{%
    }%
    \gplgaddtomacro\gplfronttext{%
    }%
    \gplgaddtomacro\gplbacktext{%
    }%
    \gplgaddtomacro\gplfronttext{%
    }%
    \gplgaddtomacro\gplbacktext{%
      \csname LTb\endcsname%
      \put(907,778){\makebox(0,0){\strut{}$0$}}%
      \csname LTb\endcsname%
      \put(3119,778){\makebox(0,0){\strut{}$500$}}%
      \csname LTb\endcsname%
      \put(5330,778){\makebox(0,0){\strut{}$1000$}}%
      \csname LTb\endcsname%
      \put(7542,778){\makebox(0,0){\strut{}$1500$}}%
      \csname LTb\endcsname%
      \put(9753,778){\makebox(0,0){\strut{}$2000$}}%
      \csname LTb\endcsname%
      \put(11965,778){\makebox(0,0){\strut{}$2500$}}%
    }%
    \gplgaddtomacro\gplfronttext{%
      \csname LTb\endcsname%
      \put(6803,499){\makebox(0,0){\strut{}Pulse number}}%
    }%
    \gplbacktext
    \put(0,0){\includegraphics{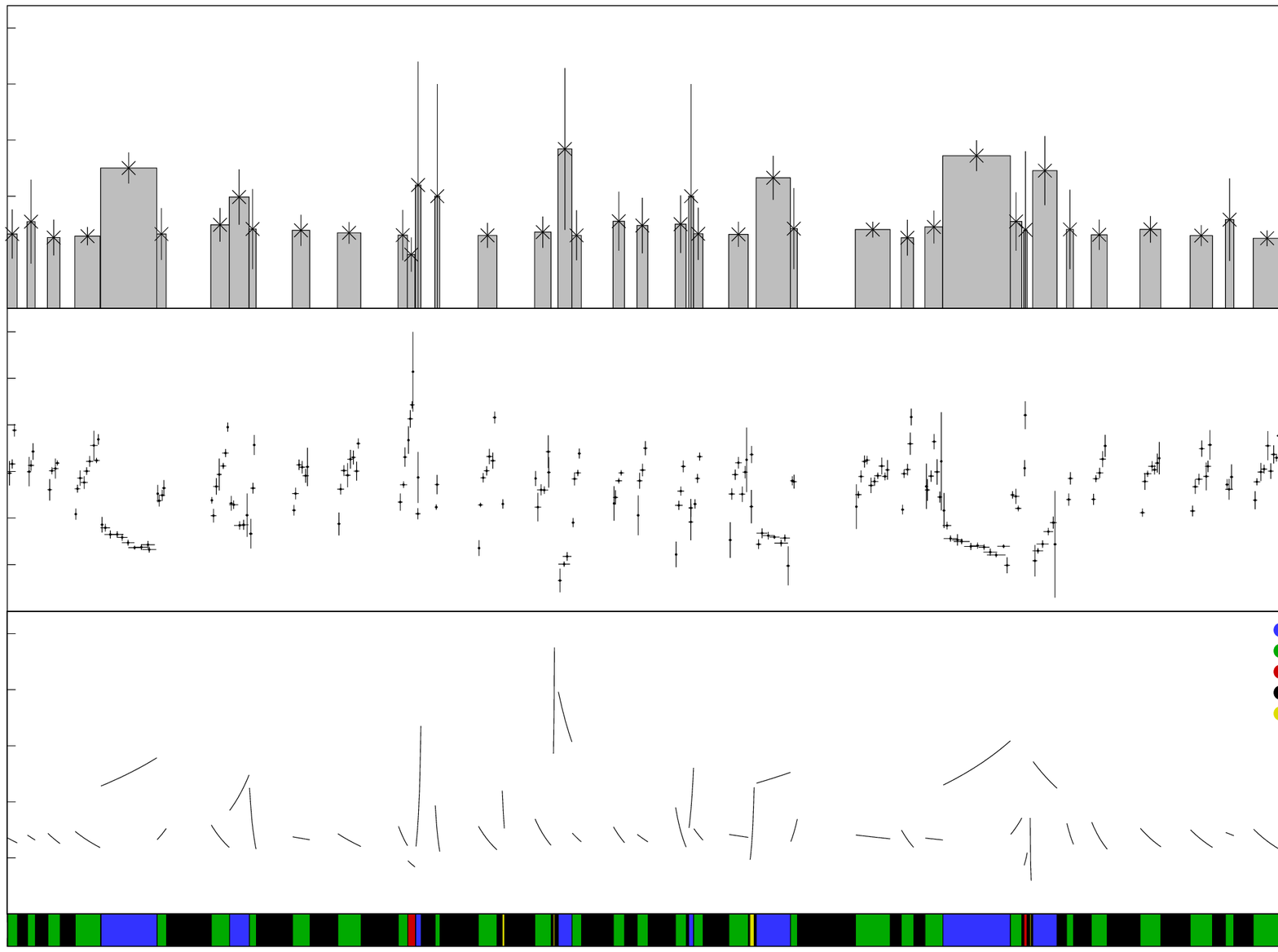}}%
    \gplfronttext
  \end{picture}%
\endgroup

%% file: all_lines.tex
% GNUPLOT: LaTeX picture with Postscript
\begingroup
  \makeatletter
  \providecommand\color[2][]{%
    \GenericError{(gnuplot) \space\space\space\@spaces}{%
      Package color not loaded in conjunction with
      terminal option `colourtext'%
    }{See the gnuplot documentation for explanation.%
    }{Either use 'blacktext' in gnuplot or load the package
      color.sty in LaTeX.}%
    \renewcommand\color[2][]{}%
  }%
  \providecommand\includegraphics[2][]{%
    \GenericError{(gnuplot) \space\space\space\@spaces}{%
      Package graphicx or graphics not loaded%
    }{See the gnuplot documentation for explanation.%
    }{The gnuplot epslatex terminal needs graphicx.sty or graphics.sty.}%
    \renewcommand\includegraphics[2][]{}%
  }%
  \providecommand\rotatebox[2]{#2}%
  \@ifundefined{ifGPcolor}{%
    \newif\ifGPcolor
    \GPcolortrue
  }{}%
  \@ifundefined{ifGPblacktext}{%
    \newif\ifGPblacktext
    \GPblacktexttrue
  }{}%
  % define a \g@addto@macro without @ in the name:
  \let\gplgaddtomacro\g@addto@macro
  % define empty templates for all commands taking text:
  \gdef\gplbacktext{}%
  \gdef\gplfronttext{}%
  \makeatother
  \ifGPblacktext
    % no textcolor at all
    \def\colorrgb#1{}%
    \def\colorgray#1{}%
  \else
    % gray or color?
    \ifGPcolor
      \def\colorrgb#1{\color[rgb]{#1}}%
      \def\colorgray#1{\color[gray]{#1}}%
      \expandafter\def\csname LTw\endcsname{\color{white}}%
      \expandafter\def\csname LTb\endcsname{\color{black}}%
      \expandafter\def\csname LTa\endcsname{\color{black}}%
      \expandafter\def\csname LT0\endcsname{\color[rgb]{1,0,0}}%
      \expandafter\def\csname LT1\endcsname{\color[rgb]{0,1,0}}%
      \expandafter\def\csname LT2\endcsname{\color[rgb]{0,0,1}}%
      \expandafter\def\csname LT3\endcsname{\color[rgb]{1,0,1}}%
      \expandafter\def\csname LT4\endcsname{\color[rgb]{0,1,1}}%
      \expandafter\def\csname LT5\endcsname{\color[rgb]{1,1,0}}%
      \expandafter\def\csname LT6\endcsname{\color[rgb]{0,0,0}}%
      \expandafter\def\csname LT7\endcsname{\color[rgb]{1,0.3,0}}%
      \expandafter\def\csname LT8\endcsname{\color[rgb]{0.5,0.5,0.5}}%
    \else
      % gray
      \def\colorrgb#1{\color{black}}%
      \def\colorgray#1{\color[gray]{#1}}%
      \expandafter\def\csname LTw\endcsname{\color{white}}%
      \expandafter\def\csname LTb\endcsname{\color{black}}%
      \expandafter\def\csname LTa\endcsname{\color{black}}%
      \expandafter\def\csname LT0\endcsname{\color{black}}%
      \expandafter\def\csname LT1\endcsname{\color{black}}%
      \expandafter\def\csname LT2\endcsname{\color{black}}%
      \expandafter\def\csname LT3\endcsname{\color{black}}%
      \expandafter\def\csname LT4\endcsname{\color{black}}%
      \expandafter\def\csname LT5\endcsname{\color{black}}%
      \expandafter\def\csname LT6\endcsname{\color{black}}%
      \expandafter\def\csname LT7\endcsname{\color{black}}%
      \expandafter\def\csname LT8\endcsname{\color{black}}%
    \fi
  \fi
    \setlength{\unitlength}{0.0500bp}%
    \ifx\gptboxheight\undefined%
      \newlength{\gptboxheight}%
      \newlength{\gptboxwidth}%
      \newsavebox{\gptboxtext}%
    \fi%
    \setlength{\fboxrule}{0.5pt}%
    \setlength{\fboxsep}{1pt}%
\begin{picture}(4320.00,6480.00)%
    \gplgaddtomacro\gplbacktext{%
      \csname LTb\endcsname%
      \put(675,611){\makebox(0,0)[r]{\strut{}$2000$}}%
      \csname LTb\endcsname%
      \put(675,1740){\makebox(0,0)[r]{\strut{}$2020$}}%
      \csname LTb\endcsname%
      \put(675,2869){\makebox(0,0)[r]{\strut{}$2040$}}%
      \csname LTb\endcsname%
      \put(675,3998){\makebox(0,0)[r]{\strut{}$2060$}}%
      \csname LTb\endcsname%
      \put(675,5127){\makebox(0,0)[r]{\strut{}$2080$}}%
      \csname LTb\endcsname%
      \put(675,6256){\makebox(0,0)[r]{\strut{}$2100$}}%
      \csname LTb\endcsname%
      \put(1080,397){\makebox(0,0){\strut{}$-20$}}%
      \csname LTb\endcsname%
      \put(1650,397){\makebox(0,0){\strut{}$-10$}}%
      \csname LTb\endcsname%
      \put(2220,397){\makebox(0,0){\strut{}$0$}}%
      \csname LTb\endcsname%
      \put(2790,397){\makebox(0,0){\strut{}$10$}}%
      \csname LTb\endcsname%
      \put(3359,397){\makebox(0,0){\strut{}$20$}}%
      \csname LTb\endcsname%
      \put(3929,397){\makebox(0,0){\strut{}$30$}}%
    }%
    \gplgaddtomacro\gplfronttext{%
      \csname LTb\endcsname%
      \put(72,3433){\rotatebox{-270}{\makebox(0,0){\strut{}Pulse number}}}%
      \csname LTb\endcsname%
      \put(2440,118){\makebox(0,0){\strut{}Rotation phase ($^\circ$)}}%
    }%
    \gplgaddtomacro\gplbacktext{%
    }%
    \gplgaddtomacro\gplfronttext{%
    }%
    \gplgaddtomacro\gplbacktext{%
    }%
    \gplgaddtomacro\gplfronttext{%
    }%
    \gplgaddtomacro\gplbacktext{%
    }%
    \gplgaddtomacro\gplfronttext{%
    }%
    \gplgaddtomacro\gplbacktext{%
    }%
    \gplgaddtomacro\gplfronttext{%
    }%
    \gplgaddtomacro\gplbacktext{%
    }%
    \gplgaddtomacro\gplfronttext{%
    }%
    \gplgaddtomacro\gplbacktext{%
    }%
    \gplgaddtomacro\gplfronttext{%
    }%
    \gplgaddtomacro\gplbacktext{%
    }%
    \gplgaddtomacro\gplfronttext{%
    }%
    \gplgaddtomacro\gplbacktext{%
    }%
    \gplgaddtomacro\gplfronttext{%
    }%
    \gplgaddtomacro\gplbacktext{%
    }%
    \gplgaddtomacro\gplfronttext{%
    }%
    \gplgaddtomacro\gplbacktext{%
    }%
    \gplgaddtomacro\gplfronttext{%
    }%
    \gplgaddtomacro\gplbacktext{%
    }%
    \gplgaddtomacro\gplfronttext{%
    }%
    \gplgaddtomacro\gplbacktext{%
    }%
    \gplgaddtomacro\gplfronttext{%
    }%
    \gplgaddtomacro\gplbacktext{%
    }%
    \gplgaddtomacro\gplfronttext{%
    }%
    \gplgaddtomacro\gplbacktext{%
    }%
    \gplgaddtomacro\gplfronttext{%
    }%
    \gplgaddtomacro\gplbacktext{%
    }%
    \gplgaddtomacro\gplfronttext{%
    }%
    \gplgaddtomacro\gplbacktext{%
    }%
    \gplgaddtomacro\gplfronttext{%
    }%
    \gplgaddtomacro\gplbacktext{%
    }%
    \gplgaddtomacro\gplfronttext{%
    }%
    \gplgaddtomacro\gplbacktext{%
    }%
    \gplgaddtomacro\gplfronttext{%
    }%
    \gplgaddtomacro\gplbacktext{%
    }%
    \gplgaddtomacro\gplfronttext{%
    }%
    \gplgaddtomacro\gplbacktext{%
    }%
    \gplgaddtomacro\gplfronttext{%
    }%
    \gplgaddtomacro\gplbacktext{%
    }%
    \gplgaddtomacro\gplfronttext{%
    }%
    \gplgaddtomacro\gplbacktext{%
    }%
    \gplgaddtomacro\gplfronttext{%
    }%
    \gplgaddtomacro\gplbacktext{%
    }%
    \gplgaddtomacro\gplfronttext{%
    }%
    \gplgaddtomacro\gplbacktext{%
    }%
    \gplgaddtomacro\gplfronttext{%
    }%
    \gplgaddtomacro\gplbacktext{%
    }%
    \gplgaddtomacro\gplfronttext{%
    }%
    \gplgaddtomacro\gplbacktext{%
    }%
    \gplgaddtomacro\gplfronttext{%
    }%
    \gplgaddtomacro\gplbacktext{%
    }%
    \gplgaddtomacro\gplfronttext{%
    }%
    \gplgaddtomacro\gplbacktext{%
    }%
    \gplgaddtomacro\gplfronttext{%
    }%
    \gplgaddtomacro\gplbacktext{%
    }%
    \gplgaddtomacro\gplfronttext{%
    }%
    \gplgaddtomacro\gplbacktext{%
    }%
    \gplgaddtomacro\gplfronttext{%
    }%
    \gplgaddtomacro\gplbacktext{%
    }%
    \gplgaddtomacro\gplfronttext{%
    }%
    \gplgaddtomacro\gplbacktext{%
    }%
    \gplgaddtomacro\gplfronttext{%
    }%
    \gplgaddtomacro\gplbacktext{%
    }%
    \gplgaddtomacro\gplfronttext{%
    }%
    \gplgaddtomacro\gplbacktext{%
    }%
    \gplgaddtomacro\gplfronttext{%
    }%
    \gplgaddtomacro\gplbacktext{%
    }%
    \gplgaddtomacro\gplfronttext{%
    }%
    \gplgaddtomacro\gplbacktext{%
    }%
    \gplgaddtomacro\gplfronttext{%
    }%
    \gplgaddtomacro\gplbacktext{%
    }%
    \gplgaddtomacro\gplfronttext{%
    }%
    \gplgaddtomacro\gplbacktext{%
    }%
    \gplgaddtomacro\gplfronttext{%
    }%
    \gplgaddtomacro\gplbacktext{%
    }%
    \gplgaddtomacro\gplfronttext{%
    }%
    \gplgaddtomacro\gplbacktext{%
    }%
    \gplgaddtomacro\gplfronttext{%
    }%
    \gplgaddtomacro\gplbacktext{%
    }%
    \gplgaddtomacro\gplfronttext{%
    }%
    \gplgaddtomacro\gplbacktext{%
    }%
    \gplgaddtomacro\gplfronttext{%
    }%
    \gplgaddtomacro\gplbacktext{%
    }%
    \gplgaddtomacro\gplfronttext{%
    }%
    \gplgaddtomacro\gplbacktext{%
    }%
    \gplgaddtomacro\gplfronttext{%
    }%
    \gplgaddtomacro\gplbacktext{%
    }%
    \gplgaddtomacro\gplfronttext{%
    }%
    \gplgaddtomacro\gplbacktext{%
    }%
    \gplgaddtomacro\gplfronttext{%
    }%
    \gplgaddtomacro\gplbacktext{%
    }%
    \gplgaddtomacro\gplfronttext{%
    }%
    \gplgaddtomacro\gplbacktext{%
    }%
    \gplgaddtomacro\gplfronttext{%
    }%
    \gplgaddtomacro\gplbacktext{%
    }%
    \gplgaddtomacro\gplfronttext{%
    }%
    \gplgaddtomacro\gplbacktext{%
    }%
    \gplgaddtomacro\gplfronttext{%
    }%
    \gplgaddtomacro\gplbacktext{%
    }%
    \gplgaddtomacro\gplfronttext{%
    }%
    \gplgaddtomacro\gplbacktext{%
    }%
    \gplgaddtomacro\gplfronttext{%
    }%
    \gplgaddtomacro\gplbacktext{%
    }%
    \gplgaddtomacro\gplfronttext{%
    }%
    \gplgaddtomacro\gplbacktext{%
    }%
    \gplgaddtomacro\gplfronttext{%
    }%
    \gplgaddtomacro\gplbacktext{%
    }%
    \gplgaddtomacro\gplfronttext{%
    }%
    \gplgaddtomacro\gplbacktext{%
    }%
    \gplgaddtomacro\gplfronttext{%
    }%
    \gplgaddtomacro\gplbacktext{%
    }%
    \gplgaddtomacro\gplfronttext{%
    }%
    \gplgaddtomacro\gplbacktext{%
    }%
    \gplgaddtomacro\gplfronttext{%
    }%
    \gplgaddtomacro\gplbacktext{%
    }%
    \gplgaddtomacro\gplfronttext{%
    }%
    \gplgaddtomacro\gplbacktext{%
    }%
    \gplgaddtomacro\gplfronttext{%
    }%
    \gplgaddtomacro\gplbacktext{%
    }%
    \gplgaddtomacro\gplfronttext{%
    }%
    \gplgaddtomacro\gplbacktext{%
    }%
    \gplgaddtomacro\gplfronttext{%
    }%
    \gplgaddtomacro\gplbacktext{%
    }%
    \gplgaddtomacro\gplfronttext{%
    }%
    \gplgaddtomacro\gplbacktext{%
    }%
    \gplgaddtomacro\gplfronttext{%
    }%
    \gplgaddtomacro\gplbacktext{%
    }%
    \gplgaddtomacro\gplfronttext{%
    }%
    \gplgaddtomacro\gplbacktext{%
    }%
    \gplgaddtomacro\gplfronttext{%
    }%
    \gplgaddtomacro\gplbacktext{%
    }%
    \gplgaddtomacro\gplfronttext{%
    }%
    \gplgaddtomacro\gplbacktext{%
    }%
    \gplgaddtomacro\gplfronttext{%
    }%
    \gplgaddtomacro\gplbacktext{%
    }%
    \gplgaddtomacro\gplfronttext{%
    }%
    \gplgaddtomacro\gplbacktext{%
    }%
    \gplgaddtomacro\gplfronttext{%
    }%
    \gplgaddtomacro\gplbacktext{%
    }%
    \gplgaddtomacro\gplfronttext{%
    }%
    \gplgaddtomacro\gplbacktext{%
    }%
    \gplgaddtomacro\gplfronttext{%
    }%
    \gplgaddtomacro\gplbacktext{%
    }%
    \gplgaddtomacro\gplfronttext{%
    }%
    \gplgaddtomacro\gplbacktext{%
    }%
    \gplgaddtomacro\gplfronttext{%
    }%
    \gplgaddtomacro\gplbacktext{%
    }%
    \gplgaddtomacro\gplfronttext{%
    }%
    \gplgaddtomacro\gplbacktext{%
    }%
    \gplgaddtomacro\gplfronttext{%
    }%
    \gplgaddtomacro\gplbacktext{%
    }%
    \gplgaddtomacro\gplfronttext{%
    }%
    \gplgaddtomacro\gplbacktext{%
    }%
    \gplgaddtomacro\gplfronttext{%
    }%
    \gplgaddtomacro\gplbacktext{%
    }%
    \gplgaddtomacro\gplfronttext{%
    }%
    \gplgaddtomacro\gplbacktext{%
    }%
    \gplgaddtomacro\gplfronttext{%
    }%
    \gplgaddtomacro\gplbacktext{%
    }%
    \gplgaddtomacro\gplfronttext{%
    }%
    \gplgaddtomacro\gplbacktext{%
    }%
    \gplgaddtomacro\gplfronttext{%
    }%
    \gplgaddtomacro\gplbacktext{%
    }%
    \gplgaddtomacro\gplfronttext{%
    }%
    \gplgaddtomacro\gplbacktext{%
    }%
    \gplgaddtomacro\gplfronttext{%
    }%
    \gplgaddtomacro\gplbacktext{%
    }%
    \gplgaddtomacro\gplfronttext{%
    }%
    \gplgaddtomacro\gplbacktext{%
    }%
    \gplgaddtomacro\gplfronttext{%
    }%
    \gplgaddtomacro\gplbacktext{%
    }%
    \gplgaddtomacro\gplfronttext{%
    }%
    \gplgaddtomacro\gplbacktext{%
    }%
    \gplgaddtomacro\gplfronttext{%
    }%
    \gplgaddtomacro\gplbacktext{%
    }%
    \gplgaddtomacro\gplfronttext{%
    }%
    \gplgaddtomacro\gplbacktext{%
    }%
    \gplgaddtomacro\gplfronttext{%
    }%
    \gplgaddtomacro\gplbacktext{%
    }%
    \gplgaddtomacro\gplfronttext{%
    }%
    \gplgaddtomacro\gplbacktext{%
    }%
    \gplgaddtomacro\gplfronttext{%
    }%
    \gplgaddtomacro\gplbacktext{%
    }%
    \gplgaddtomacro\gplfronttext{%
    }%
    \gplgaddtomacro\gplbacktext{%
    }%
    \gplgaddtomacro\gplfronttext{%
    }%
    \gplgaddtomacro\gplbacktext{%
    }%
    \gplgaddtomacro\gplfronttext{%
    }%
    \gplgaddtomacro\gplbacktext{%
    }%
    \gplgaddtomacro\gplfronttext{%
    }%
    \gplgaddtomacro\gplbacktext{%
    }%
    \gplgaddtomacro\gplfronttext{%
    }%
    \gplgaddtomacro\gplbacktext{%
    }%
    \gplgaddtomacro\gplfronttext{%
    }%
    \gplgaddtomacro\gplbacktext{%
    }%
    \gplgaddtomacro\gplfronttext{%
    }%
    \gplgaddtomacro\gplbacktext{%
    }%
    \gplgaddtomacro\gplfronttext{%
    }%
    \gplgaddtomacro\gplbacktext{%
    }%
    \gplgaddtomacro\gplfronttext{%
    }%
    \gplgaddtomacro\gplbacktext{%
    }%
    \gplgaddtomacro\gplfronttext{%
    }%
    \gplgaddtomacro\gplbacktext{%
    }%
    \gplgaddtomacro\gplfronttext{%
    }%
    \gplgaddtomacro\gplbacktext{%
    }%
    \gplgaddtomacro\gplfronttext{%
    }%
    \gplgaddtomacro\gplbacktext{%
    }%
    \gplgaddtomacro\gplfronttext{%
    }%
    \gplgaddtomacro\gplbacktext{%
    }%
    \gplgaddtomacro\gplfronttext{%
    }%
    \gplgaddtomacro\gplbacktext{%
    }%
    \gplgaddtomacro\gplfronttext{%
    }%
    \gplgaddtomacro\gplbacktext{%
    }%
    \gplgaddtomacro\gplfronttext{%
    }%
    \gplgaddtomacro\gplbacktext{%
    }%
    \gplgaddtomacro\gplfronttext{%
    }%
    \gplgaddtomacro\gplbacktext{%
    }%
    \gplgaddtomacro\gplfronttext{%
    }%
    \gplgaddtomacro\gplbacktext{%
    }%
    \gplgaddtomacro\gplfronttext{%
    }%
    \gplgaddtomacro\gplbacktext{%
    }%
    \gplgaddtomacro\gplfronttext{%
    }%
    \gplgaddtomacro\gplbacktext{%
    }%
    \gplgaddtomacro\gplfronttext{%
    }%
    \gplgaddtomacro\gplbacktext{%
    }%
    \gplgaddtomacro\gplfronttext{%
    }%
    \gplgaddtomacro\gplbacktext{%
    }%
    \gplgaddtomacro\gplfronttext{%
    }%
    \gplgaddtomacro\gplbacktext{%
    }%
    \gplgaddtomacro\gplfronttext{%
    }%
    \gplgaddtomacro\gplbacktext{%
    }%
    \gplgaddtomacro\gplfronttext{%
    }%
    \gplgaddtomacro\gplbacktext{%
    }%
    \gplgaddtomacro\gplfronttext{%
    }%
    \gplgaddtomacro\gplbacktext{%
    }%
    \gplgaddtomacro\gplfronttext{%
    }%
    \gplgaddtomacro\gplbacktext{%
    }%
    \gplgaddtomacro\gplfronttext{%
    }%
    \gplgaddtomacro\gplbacktext{%
    }%
    \gplgaddtomacro\gplfronttext{%
    }%
    \gplgaddtomacro\gplbacktext{%
    }%
    \gplgaddtomacro\gplfronttext{%
    }%
    \gplgaddtomacro\gplbacktext{%
    }%
    \gplgaddtomacro\gplfronttext{%
    }%
    \gplgaddtomacro\gplbacktext{%
    }%
    \gplgaddtomacro\gplfronttext{%
    }%
    \gplgaddtomacro\gplbacktext{%
    }%
    \gplgaddtomacro\gplfronttext{%
    }%
    \gplgaddtomacro\gplbacktext{%
    }%
    \gplgaddtomacro\gplfronttext{%
    }%
    \gplgaddtomacro\gplbacktext{%
    }%
    \gplgaddtomacro\gplfronttext{%
    }%
    \gplgaddtomacro\gplbacktext{%
    }%
    \gplgaddtomacro\gplfronttext{%
    }%
    \gplgaddtomacro\gplbacktext{%
    }%
    \gplgaddtomacro\gplfronttext{%
    }%
    \gplgaddtomacro\gplbacktext{%
    }%
    \gplgaddtomacro\gplfronttext{%
    }%
    \gplgaddtomacro\gplbacktext{%
    }%
    \gplgaddtomacro\gplfronttext{%
    }%
    \gplgaddtomacro\gplbacktext{%
    }%
    \gplgaddtomacro\gplfronttext{%
    }%
    \gplgaddtomacro\gplbacktext{%
    }%
    \gplgaddtomacro\gplfronttext{%
    }%
    \gplgaddtomacro\gplbacktext{%
    }%
    \gplgaddtomacro\gplfronttext{%
    }%
    \gplgaddtomacro\gplbacktext{%
    }%
    \gplgaddtomacro\gplfronttext{%
    }%
    \gplgaddtomacro\gplbacktext{%
    }%
    \gplgaddtomacro\gplfronttext{%
    }%
    \gplgaddtomacro\gplbacktext{%
    }%
    \gplgaddtomacro\gplfronttext{%
    }%
    \gplgaddtomacro\gplbacktext{%
    }%
    \gplgaddtomacro\gplfronttext{%
    }%
    \gplgaddtomacro\gplbacktext{%
    }%
    \gplgaddtomacro\gplfronttext{%
    }%
    \gplgaddtomacro\gplbacktext{%
    }%
    \gplgaddtomacro\gplfronttext{%
    }%
    \gplgaddtomacro\gplbacktext{%
    }%
    \gplgaddtomacro\gplfronttext{%
    }%
    \gplgaddtomacro\gplbacktext{%
    }%
    \gplgaddtomacro\gplfronttext{%
    }%
    \gplgaddtomacro\gplbacktext{%
    }%
    \gplgaddtomacro\gplfronttext{%
    }%
    \gplgaddtomacro\gplbacktext{%
    }%
    \gplgaddtomacro\gplfronttext{%
    }%
    \gplgaddtomacro\gplbacktext{%
    }%
    \gplgaddtomacro\gplfronttext{%
    }%
    \gplgaddtomacro\gplbacktext{%
    }%
    \gplgaddtomacro\gplfronttext{%
    }%
    \gplgaddtomacro\gplbacktext{%
    }%
    \gplgaddtomacro\gplfronttext{%
    }%
    \gplgaddtomacro\gplbacktext{%
    }%
    \gplgaddtomacro\gplfronttext{%
    }%
    \gplgaddtomacro\gplbacktext{%
    }%
    \gplgaddtomacro\gplfronttext{%
    }%
    \gplgaddtomacro\gplbacktext{%
    }%
    \gplgaddtomacro\gplfronttext{%
    }%
    \gplgaddtomacro\gplbacktext{%
    }%
    \gplgaddtomacro\gplfronttext{%
    }%
    \gplgaddtomacro\gplbacktext{%
    }%
    \gplgaddtomacro\gplfronttext{%
    }%
    \gplgaddtomacro\gplbacktext{%
    }%
    \gplgaddtomacro\gplfronttext{%
    }%
    \gplgaddtomacro\gplbacktext{%
    }%
    \gplgaddtomacro\gplfronttext{%
    }%
    \gplgaddtomacro\gplbacktext{%
    }%
    \gplgaddtomacro\gplfronttext{%
    }%
    \gplgaddtomacro\gplbacktext{%
    }%
    \gplgaddtomacro\gplfronttext{%
    }%
    \gplgaddtomacro\gplbacktext{%
    }%
    \gplgaddtomacro\gplfronttext{%
    }%
    \gplgaddtomacro\gplbacktext{%
    }%
    \gplgaddtomacro\gplfronttext{%
    }%
    \gplgaddtomacro\gplbacktext{%
    }%
    \gplgaddtomacro\gplfronttext{%
    }%
    \gplgaddtomacro\gplbacktext{%
    }%
    \gplgaddtomacro\gplfronttext{%
    }%
    \gplgaddtomacro\gplbacktext{%
    }%
    \gplgaddtomacro\gplfronttext{%
    }%
    \gplgaddtomacro\gplbacktext{%
    }%
    \gplgaddtomacro\gplfronttext{%
    }%
    \gplgaddtomacro\gplbacktext{%
    }%
    \gplgaddtomacro\gplfronttext{%
    }%
    \gplgaddtomacro\gplbacktext{%
    }%
    \gplgaddtomacro\gplfronttext{%
    }%
    \gplgaddtomacro\gplbacktext{%
    }%
    \gplgaddtomacro\gplfronttext{%
    }%
    \gplgaddtomacro\gplbacktext{%
    }%
    \gplgaddtomacro\gplfronttext{%
    }%
    \gplgaddtomacro\gplbacktext{%
    }%
    \gplgaddtomacro\gplfronttext{%
    }%
    \gplgaddtomacro\gplbacktext{%
    }%
    \gplgaddtomacro\gplfronttext{%
    }%
    \gplgaddtomacro\gplbacktext{%
    }%
    \gplgaddtomacro\gplfronttext{%
    }%
    \gplgaddtomacro\gplbacktext{%
    }%
    \gplgaddtomacro\gplfronttext{%
    }%
    \gplgaddtomacro\gplbacktext{%
    }%
    \gplgaddtomacro\gplfronttext{%
    }%
    \gplgaddtomacro\gplbacktext{%
    }%
    \gplgaddtomacro\gplfronttext{%
    }%
    \gplgaddtomacro\gplbacktext{%
    }%
    \gplgaddtomacro\gplfronttext{%
    }%
    \gplgaddtomacro\gplbacktext{%
    }%
    \gplgaddtomacro\gplfronttext{%
    }%
    \gplgaddtomacro\gplbacktext{%
    }%
    \gplgaddtomacro\gplfronttext{%
    }%
    \gplgaddtomacro\gplbacktext{%
    }%
    \gplgaddtomacro\gplfronttext{%
    }%
    \gplgaddtomacro\gplbacktext{%
    }%
    \gplgaddtomacro\gplfronttext{%
    }%
    \gplgaddtomacro\gplbacktext{%
    }%
    \gplgaddtomacro\gplfronttext{%
    }%
    \gplgaddtomacro\gplbacktext{%
    }%
    \gplgaddtomacro\gplfronttext{%
    }%
    \gplgaddtomacro\gplbacktext{%
    }%
    \gplgaddtomacro\gplfronttext{%
    }%
    \gplgaddtomacro\gplbacktext{%
    }%
    \gplgaddtomacro\gplfronttext{%
    }%
    \gplgaddtomacro\gplbacktext{%
    }%
    \gplgaddtomacro\gplfronttext{%
    }%
    \gplgaddtomacro\gplbacktext{%
    }%
    \gplgaddtomacro\gplfronttext{%
    }%
    \gplgaddtomacro\gplbacktext{%
    }%
    \gplgaddtomacro\gplfronttext{%
    }%
    \gplgaddtomacro\gplbacktext{%
    }%
    \gplgaddtomacro\gplfronttext{%
    }%
    \gplgaddtomacro\gplbacktext{%
    }%
    \gplgaddtomacro\gplfronttext{%
    }%
    \gplgaddtomacro\gplbacktext{%
    }%
    \gplgaddtomacro\gplfronttext{%
    }%
    \gplgaddtomacro\gplbacktext{%
    }%
    \gplgaddtomacro\gplfronttext{%
    }%
    \gplgaddtomacro\gplbacktext{%
    }%
    \gplgaddtomacro\gplfronttext{%
    }%
    \gplgaddtomacro\gplbacktext{%
    }%
    \gplgaddtomacro\gplfronttext{%
    }%
    \gplgaddtomacro\gplbacktext{%
    }%
    \gplgaddtomacro\gplfronttext{%
    }%
    \gplgaddtomacro\gplbacktext{%
    }%
    \gplgaddtomacro\gplfronttext{%
    }%
    \gplgaddtomacro\gplbacktext{%
    }%
    \gplgaddtomacro\gplfronttext{%
    }%
    \gplgaddtomacro\gplbacktext{%
    }%
    \gplgaddtomacro\gplfronttext{%
    }%
    \gplgaddtomacro\gplbacktext{%
    }%
    \gplgaddtomacro\gplfronttext{%
    }%
    \gplgaddtomacro\gplbacktext{%
    }%
    \gplgaddtomacro\gplfronttext{%
    }%
    \gplgaddtomacro\gplbacktext{%
    }%
    \gplgaddtomacro\gplfronttext{%
    }%
    \gplgaddtomacro\gplbacktext{%
    }%
    \gplgaddtomacro\gplfronttext{%
    }%
    \gplgaddtomacro\gplbacktext{%
    }%
    \gplgaddtomacro\gplfronttext{%
    }%
    \gplgaddtomacro\gplbacktext{%
    }%
    \gplgaddtomacro\gplfronttext{%
    }%
    \gplgaddtomacro\gplbacktext{%
    }%
    \gplgaddtomacro\gplfronttext{%
    }%
    \gplgaddtomacro\gplbacktext{%
    }%
    \gplgaddtomacro\gplfronttext{%
    }%
    \gplgaddtomacro\gplbacktext{%
    }%
    \gplgaddtomacro\gplfronttext{%
    }%
    \gplgaddtomacro\gplbacktext{%
    }%
    \gplgaddtomacro\gplfronttext{%
    }%
    \gplgaddtomacro\gplbacktext{%
    }%
    \gplgaddtomacro\gplfronttext{%
    }%
    \gplgaddtomacro\gplbacktext{%
    }%
    \gplgaddtomacro\gplfronttext{%
    }%
    \gplgaddtomacro\gplbacktext{%
    }%
    \gplgaddtomacro\gplfronttext{%
    }%
    \gplgaddtomacro\gplbacktext{%
    }%
    \gplgaddtomacro\gplfronttext{%
    }%
    \gplgaddtomacro\gplbacktext{%
    }%
    \gplgaddtomacro\gplfronttext{%
    }%
    \gplgaddtomacro\gplbacktext{%
    }%
    \gplgaddtomacro\gplfronttext{%
    }%
    \gplgaddtomacro\gplbacktext{%
    }%
    \gplgaddtomacro\gplfronttext{%
    }%
    \gplgaddtomacro\gplbacktext{%
    }%
    \gplgaddtomacro\gplfronttext{%
    }%
    \gplgaddtomacro\gplbacktext{%
    }%
    \gplgaddtomacro\gplfronttext{%
    }%
    \gplgaddtomacro\gplbacktext{%
    }%
    \gplgaddtomacro\gplfronttext{%
    }%
    \gplgaddtomacro\gplbacktext{%
    }%
    \gplgaddtomacro\gplfronttext{%
    }%
    \gplgaddtomacro\gplbacktext{%
    }%
    \gplgaddtomacro\gplfronttext{%
    }%
    \gplgaddtomacro\gplbacktext{%
    }%
    \gplgaddtomacro\gplfronttext{%
    }%
    \gplgaddtomacro\gplbacktext{%
    }%
    \gplgaddtomacro\gplfronttext{%
    }%
    \gplgaddtomacro\gplbacktext{%
    }%
    \gplgaddtomacro\gplfronttext{%
    }%
    \gplgaddtomacro\gplbacktext{%
    }%
    \gplgaddtomacro\gplfronttext{%
    }%
    \gplgaddtomacro\gplbacktext{%
    }%
    \gplgaddtomacro\gplfronttext{%
    }%
    \gplgaddtomacro\gplbacktext{%
    }%
    \gplgaddtomacro\gplfronttext{%
    }%
    \gplgaddtomacro\gplbacktext{%
    }%
    \gplgaddtomacro\gplfronttext{%
    }%
    \gplgaddtomacro\gplbacktext{%
    }%
    \gplgaddtomacro\gplfronttext{%
    }%
    \gplgaddtomacro\gplbacktext{%
    }%
    \gplgaddtomacro\gplfronttext{%
    }%
    \gplgaddtomacro\gplbacktext{%
    }%
    \gplgaddtomacro\gplfronttext{%
    }%
    \gplgaddtomacro\gplbacktext{%
    }%
    \gplgaddtomacro\gplfronttext{%
    }%
    \gplgaddtomacro\gplbacktext{%
    }%
    \gplgaddtomacro\gplfronttext{%
    }%
    \gplgaddtomacro\gplbacktext{%
    }%
    \gplgaddtomacro\gplfronttext{%
    }%
    \gplgaddtomacro\gplbacktext{%
    }%
    \gplgaddtomacro\gplfronttext{%
    }%
    \gplgaddtomacro\gplbacktext{%
    }%
    \gplgaddtomacro\gplfronttext{%
    }%
    \gplgaddtomacro\gplbacktext{%
    }%
    \gplgaddtomacro\gplfronttext{%
    }%
    \gplgaddtomacro\gplbacktext{%
    }%
    \gplgaddtomacro\gplfronttext{%
    }%
    \gplgaddtomacro\gplbacktext{%
    }%
    \gplgaddtomacro\gplfronttext{%
    }%
    \gplgaddtomacro\gplbacktext{%
    }%
    \gplgaddtomacro\gplfronttext{%
    }%
    \gplgaddtomacro\gplbacktext{%
    }%
    \gplgaddtomacro\gplfronttext{%
    }%
    \gplgaddtomacro\gplbacktext{%
    }%
    \gplgaddtomacro\gplfronttext{%
    }%
    \gplgaddtomacro\gplbacktext{%
    }%
    \gplgaddtomacro\gplfronttext{%
    }%
    \gplgaddtomacro\gplbacktext{%
    }%
    \gplgaddtomacro\gplfronttext{%
    }%
    \gplgaddtomacro\gplbacktext{%
    }%
    \gplgaddtomacro\gplfronttext{%
    }%
    \gplgaddtomacro\gplbacktext{%
    }%
    \gplgaddtomacro\gplfronttext{%
    }%
    \gplgaddtomacro\gplbacktext{%
    }%
    \gplgaddtomacro\gplfronttext{%
    }%
    \gplgaddtomacro\gplbacktext{%
    }%
    \gplgaddtomacro\gplfronttext{%
    }%
    \gplgaddtomacro\gplbacktext{%
    }%
    \gplgaddtomacro\gplfronttext{%
    }%
    \gplgaddtomacro\gplbacktext{%
    }%
    \gplgaddtomacro\gplfronttext{%
    }%
    \gplgaddtomacro\gplbacktext{%
    }%
    \gplgaddtomacro\gplfronttext{%
    }%
    \gplgaddtomacro\gplbacktext{%
    }%
    \gplgaddtomacro\gplfronttext{%
    }%
    \gplgaddtomacro\gplbacktext{%
    }%
    \gplgaddtomacro\gplfronttext{%
    }%
    \gplgaddtomacro\gplbacktext{%
    }%
    \gplgaddtomacro\gplfronttext{%
    }%
    \gplgaddtomacro\gplbacktext{%
    }%
    \gplgaddtomacro\gplfronttext{%
    }%
    \gplgaddtomacro\gplbacktext{%
    }%
    \gplgaddtomacro\gplfronttext{%
    }%
    \gplgaddtomacro\gplbacktext{%
    }%
    \gplgaddtomacro\gplfronttext{%
    }%
    \gplgaddtomacro\gplbacktext{%
    }%
    \gplgaddtomacro\gplfronttext{%
    }%
    \gplgaddtomacro\gplbacktext{%
    }%
    \gplgaddtomacro\gplfronttext{%
    }%
    \gplgaddtomacro\gplbacktext{%
    }%
    \gplgaddtomacro\gplfronttext{%
    }%
    \gplgaddtomacro\gplbacktext{%
    }%
    \gplgaddtomacro\gplfronttext{%
    }%
    \gplgaddtomacro\gplbacktext{%
    }%
    \gplgaddtomacro\gplfronttext{%
    }%
    \gplgaddtomacro\gplbacktext{%
    }%
    \gplgaddtomacro\gplfronttext{%
    }%
    \gplgaddtomacro\gplbacktext{%
    }%
    \gplgaddtomacro\gplfronttext{%
    }%
    \gplgaddtomacro\gplbacktext{%
    }%
    \gplgaddtomacro\gplfronttext{%
    }%
    \gplgaddtomacro\gplbacktext{%
    }%
    \gplgaddtomacro\gplfronttext{%
    }%
    \gplgaddtomacro\gplbacktext{%
    }%
    \gplgaddtomacro\gplfronttext{%
    }%
    \gplgaddtomacro\gplbacktext{%
    }%
    \gplgaddtomacro\gplfronttext{%
    }%
    \gplgaddtomacro\gplbacktext{%
    }%
    \gplgaddtomacro\gplfronttext{%
    }%
    \gplgaddtomacro\gplbacktext{%
    }%
    \gplgaddtomacro\gplfronttext{%
    }%
    \gplgaddtomacro\gplbacktext{%
    }%
    \gplgaddtomacro\gplfronttext{%
    }%
    \gplgaddtomacro\gplbacktext{%
    }%
    \gplgaddtomacro\gplfronttext{%
    }%
    \gplgaddtomacro\gplbacktext{%
    }%
    \gplgaddtomacro\gplfronttext{%
    }%
    \gplgaddtomacro\gplbacktext{%
    }%
    \gplgaddtomacro\gplfronttext{%
    }%
    \gplgaddtomacro\gplbacktext{%
    }%
    \gplgaddtomacro\gplfronttext{%
    }%
    \gplgaddtomacro\gplbacktext{%
    }%
    \gplgaddtomacro\gplfronttext{%
    }%
    \gplgaddtomacro\gplbacktext{%
    }%
    \gplgaddtomacro\gplfronttext{%
    }%
    \gplbacktext
    \put(0,0){\includegraphics{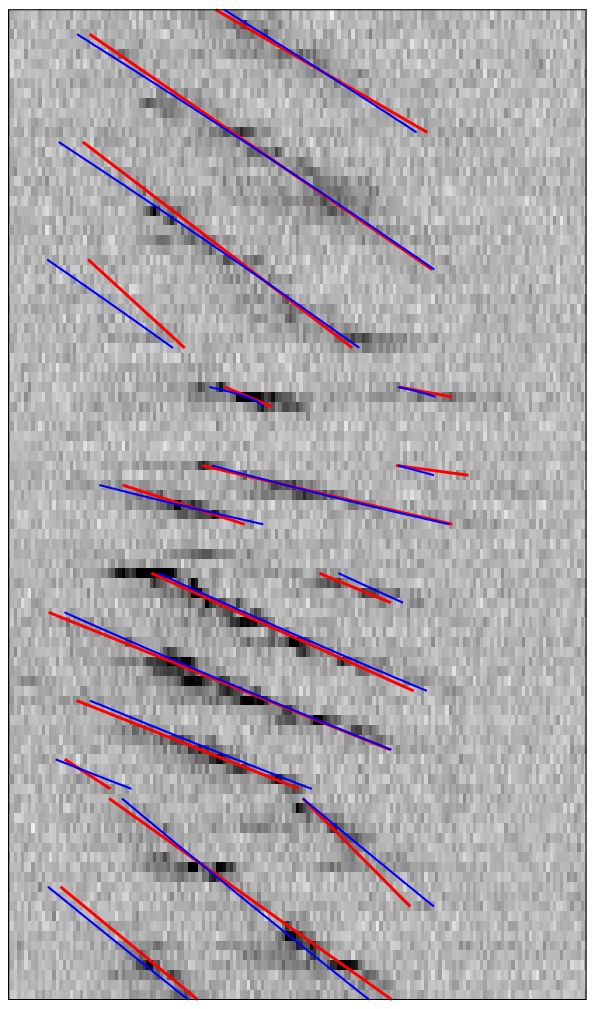}}%
    \gplfronttext
  \end{picture}%
\endgroup

%% file: P2historical.tex
% GNUPLOT: LaTeX picture with Postscript
\begingroup
  \makeatletter
  \providecommand\color[2][]{%
    \GenericError{(gnuplot) \space\space\space\@spaces}{%
      Package color not loaded in conjunction with
      terminal option `colourtext'%
    }{See the gnuplot documentation for explanation.%
    }{Either use 'blacktext' in gnuplot or load the package
      color.sty in LaTeX.}%
    \renewcommand\color[2][]{}%
  }%
  \providecommand\includegraphics[2][]{%
    \GenericError{(gnuplot) \space\space\space\@spaces}{%
      Package graphicx or graphics not loaded%
    }{See the gnuplot documentation for explanation.%
    }{The gnuplot epslatex terminal needs graphicx.sty or graphics.sty.}%
    \renewcommand\includegraphics[2][]{}%
  }%
  \providecommand\rotatebox[2]{#2}%
  \@ifundefined{ifGPcolor}{%
    \newif\ifGPcolor
    \GPcolortrue
  }{}%
  \@ifundefined{ifGPblacktext}{%
    \newif\ifGPblacktext
    \GPblacktexttrue
  }{}%
  % define a \g@addto@macro without @ in the name:
  \let\gplgaddtomacro\g@addto@macro
  % define empty templates for all commands taking text:
  \gdef\gplbacktext{}%
  \gdef\gplfronttext{}%
  \makeatother
  \ifGPblacktext
    % no textcolor at all
    \def\colorrgb#1{}%
    \def\colorgray#1{}%
  \else
    % gray or color?
    \ifGPcolor
      \def\colorrgb#1{\color[rgb]{#1}}%
      \def\colorgray#1{\color[gray]{#1}}%
      \expandafter\def\csname LTw\endcsname{\color{white}}%
      \expandafter\def\csname LTb\endcsname{\color{black}}%
      \expandafter\def\csname LTa\endcsname{\color{black}}%
      \expandafter\def\csname LT0\endcsname{\color[rgb]{1,0,0}}%
      \expandafter\def\csname LT1\endcsname{\color[rgb]{0,1,0}}%
      \expandafter\def\csname LT2\endcsname{\color[rgb]{0,0,1}}%
      \expandafter\def\csname LT3\endcsname{\color[rgb]{1,0,1}}%
      \expandafter\def\csname LT4\endcsname{\color[rgb]{0,1,1}}%
      \expandafter\def\csname LT5\endcsname{\color[rgb]{1,1,0}}%
      \expandafter\def\csname LT6\endcsname{\color[rgb]{0,0,0}}%
      \expandafter\def\csname LT7\endcsname{\color[rgb]{1,0.3,0}}%
      \expandafter\def\csname LT8\endcsname{\color[rgb]{0.5,0.5,0.5}}%
    \else
      % gray
      \def\colorrgb#1{\color{black}}%
      \def\colorgray#1{\color[gray]{#1}}%
      \expandafter\def\csname LTw\endcsname{\color{white}}%
      \expandafter\def\csname LTb\endcsname{\color{black}}%
      \expandafter\def\csname LTa\endcsname{\color{black}}%
      \expandafter\def\csname LT0\endcsname{\color{black}}%
      \expandafter\def\csname LT1\endcsname{\color{black}}%
      \expandafter\def\csname LT2\endcsname{\color{black}}%
      \expandafter\def\csname LT3\endcsname{\color{black}}%
      \expandafter\def\csname LT4\endcsname{\color{black}}%
      \expandafter\def\csname LT5\endcsname{\color{black}}%
      \expandafter\def\csname LT6\endcsname{\color{black}}%
      \expandafter\def\csname LT7\endcsname{\color{black}}%
      \expandafter\def\csname LT8\endcsname{\color{black}}%
    \fi
  \fi
    \setlength{\unitlength}{0.0500bp}%
    \ifx\gptboxheight\undefined%
      \newlength{\gptboxheight}%
      \newlength{\gptboxwidth}%
      \newsavebox{\gptboxtext}%
    \fi%
    \setlength{\fboxrule}{0.5pt}%
    \setlength{\fboxsep}{1pt}%
\begin{picture}(9640.00,3600.00)%
    \gplgaddtomacro\gplbacktext{%
      \csname LTb\endcsname%
      \put(669,720){\makebox(0,0)[r]{\strut{}$0$}}%
      \csname LTb\endcsname%
      \put(669,1296){\makebox(0,0)[r]{\strut{}$10$}}%
      \csname LTb\endcsname%
      \put(669,1872){\makebox(0,0)[r]{\strut{}$20$}}%
      \csname LTb\endcsname%
      \put(669,2447){\makebox(0,0)[r]{\strut{}$30$}}%
      \csname LTb\endcsname%
      \put(669,3023){\makebox(0,0)[r]{\strut{}$40$}}%
      \csname LTb\endcsname%
      \put(669,3599){\makebox(0,0)[r]{\strut{}$50$}}%
      \csname LTb\endcsname%
      \put(1157,534){\makebox(0,0){\strut{}$0.1$}}%
      \csname LTb\endcsname%
      \put(2439,534){\makebox(0,0){\strut{}$1$}}%
      \csname LTb\endcsname%
      \put(1258,835){\makebox(0,0)[l]{\strut{}Mode A}}%
    }%
    \gplgaddtomacro\gplfronttext{%
      \csname LTb\endcsname%
      \put(270,2159){\rotatebox{-270}{\makebox(0,0){\strut{}$P_2$ (deg)}}}%
    }%
    \gplgaddtomacro\gplbacktext{%
      \csname LTb\endcsname%
      \put(4049,534){\makebox(0,0){\strut{}$0.1$}}%
      \csname LTb\endcsname%
      \put(5331,534){\makebox(0,0){\strut{}$1$}}%
      \csname LTb\endcsname%
      \put(4150,835){\makebox(0,0)[l]{\strut{}Mode B}}%
    }%
    \gplgaddtomacro\gplfronttext{%
      \csname LTb\endcsname%
      \put(5108,255){\makebox(0,0){\strut{}Observing Frequency (GHz)}}%
    }%
    \gplgaddtomacro\gplbacktext{%
      \csname LTb\endcsname%
      \put(6941,534){\makebox(0,0){\strut{}$0.1$}}%
      \csname LTb\endcsname%
      \put(8223,534){\makebox(0,0){\strut{}$1$}}%
      \csname LTb\endcsname%
      \put(7042,835){\makebox(0,0)[l]{\strut{}Mode C}}%
    }%
    \gplgaddtomacro\gplfronttext{%
      \csname LTb\endcsname%
      \put(8658,3432){\makebox(0,0)[r]{\strut{}\tiny This work}}%
      \csname LTb\endcsname%
      \put(8658,3246){\makebox(0,0)[r]{\strut{}\tiny Karuppusamy et al. (2011)}}%
      \csname LTb\endcsname%
      \put(8658,3060){\makebox(0,0)[r]{\strut{}\tiny Weltevrede et al. (2007)}}%
      \csname LTb\endcsname%
      \put(8658,2874){\makebox(0,0)[r]{\strut{}\tiny Smits et al. (2007)}}%
      \csname LTb\endcsname%
      \put(8658,2688){\makebox(0,0)[r]{\strut{}\tiny Weltevrede et al. (2006)}}%
      \csname LTb\endcsname%
      \put(8658,2502){\makebox(0,0)[r]{\strut{}\tiny Smits et al. (2005)}}%
      \csname LTb\endcsname%
      \put(8658,2316){\makebox(0,0)[r]{\strut{}\tiny Izvekova et al. (1993)}}%
      \csname LTb\endcsname%
      \put(8658,2130){\makebox(0,0)[r]{\strut{}\tiny Huguenin et al. (1970)}}%
    }%
    \gplbacktext
    \put(0,0){\includegraphics{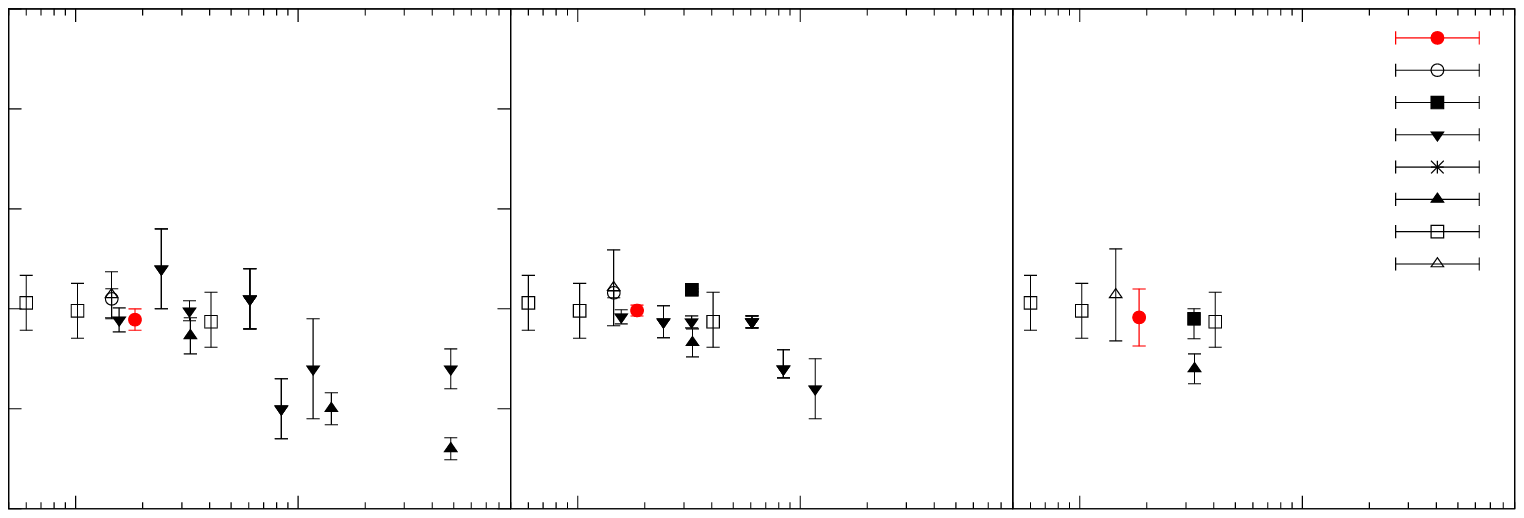}}%
    \gplfronttext
  \end{picture}%
\endgroup

%% file: residual_comparison.tex
% GNUPLOT: LaTeX picture with Postscript
\begingroup
  \makeatletter
  \providecommand\color[2][]{%
    \GenericError{(gnuplot) \space\space\space\@spaces}{%
      Package color not loaded in conjunction with
      terminal option `colourtext'%
    }{See the gnuplot documentation for explanation.%
    }{Either use 'blacktext' in gnuplot or load the package
      color.sty in LaTeX.}%
    \renewcommand\color[2][]{}%
  }%
  \providecommand\includegraphics[2][]{%
    \GenericError{(gnuplot) \space\space\space\@spaces}{%
      Package graphicx or graphics not loaded%
    }{See the gnuplot documentation for explanation.%
    }{The gnuplot epslatex terminal needs graphicx.sty or graphics.sty.}%
    \renewcommand\includegraphics[2][]{}%
  }%
  \providecommand\rotatebox[2]{#2}%
  \@ifundefined{ifGPcolor}{%
    \newif\ifGPcolor
    \GPcolortrue
  }{}%
  \@ifundefined{ifGPblacktext}{%
    \newif\ifGPblacktext
    \GPblacktexttrue
  }{}%
  % define a \g@addto@macro without @ in the name:
  \let\gplgaddtomacro\g@addto@macro
  % define empty templates for all commands taking text:
  \gdef\gplbacktext{}%
  \gdef\gplfronttext{}%
  \makeatother
  \ifGPblacktext
    % no textcolor at all
    \def\colorrgb#1{}%
    \def\colorgray#1{}%
  \else
    % gray or color?
    \ifGPcolor
      \def\colorrgb#1{\color[rgb]{#1}}%
      \def\colorgray#1{\color[gray]{#1}}%
      \expandafter\def\csname LTw\endcsname{\color{white}}%
      \expandafter\def\csname LTb\endcsname{\color{black}}%
      \expandafter\def\csname LTa\endcsname{\color{black}}%
      \expandafter\def\csname LT0\endcsname{\color[rgb]{1,0,0}}%
      \expandafter\def\csname LT1\endcsname{\color[rgb]{0,1,0}}%
      \expandafter\def\csname LT2\endcsname{\color[rgb]{0,0,1}}%
      \expandafter\def\csname LT3\endcsname{\color[rgb]{1,0,1}}%
      \expandafter\def\csname LT4\endcsname{\color[rgb]{0,1,1}}%
      \expandafter\def\csname LT5\endcsname{\color[rgb]{1,1,0}}%
      \expandafter\def\csname LT6\endcsname{\color[rgb]{0,0,0}}%
      \expandafter\def\csname LT7\endcsname{\color[rgb]{1,0.3,0}}%
      \expandafter\def\csname LT8\endcsname{\color[rgb]{0.5,0.5,0.5}}%
    \else
      % gray
      \def\colorrgb#1{\color{black}}%
      \def\colorgray#1{\color[gray]{#1}}%
      \expandafter\def\csname LTw\endcsname{\color{white}}%
      \expandafter\def\csname LTb\endcsname{\color{black}}%
      \expandafter\def\csname LTa\endcsname{\color{black}}%
      \expandafter\def\csname LT0\endcsname{\color{black}}%
      \expandafter\def\csname LT1\endcsname{\color{black}}%
      \expandafter\def\csname LT2\endcsname{\color{black}}%
      \expandafter\def\csname LT3\endcsname{\color{black}}%
      \expandafter\def\csname LT4\endcsname{\color{black}}%
      \expandafter\def\csname LT5\endcsname{\color{black}}%
      \expandafter\def\csname LT6\endcsname{\color{black}}%
      \expandafter\def\csname LT7\endcsname{\color{black}}%
      \expandafter\def\csname LT8\endcsname{\color{black}}%
    \fi
  \fi
    \setlength{\unitlength}{0.0500bp}%
    \ifx\gptboxheight\undefined%
      \newlength{\gptboxheight}%
      \newlength{\gptboxwidth}%
      \newsavebox{\gptboxtext}%
    \fi%
    \setlength{\fboxrule}{0.5pt}%
    \setlength{\fboxsep}{1pt}%
\begin{picture}(9640.00,2880.00)%
    \gplgaddtomacro\gplbacktext{%
      \csname LTb\endcsname%
      \put(862,558){\makebox(0,0)[r]{\strut{}$-15$}}%
      \csname LTb\endcsname%
      \put(862,945){\makebox(0,0)[r]{\strut{}$-10$}}%
      \csname LTb\endcsname%
      \put(862,1332){\makebox(0,0)[r]{\strut{}$-5$}}%
      \csname LTb\endcsname%
      \put(862,1719){\makebox(0,0)[r]{\strut{}$0$}}%
      \csname LTb\endcsname%
      \put(862,2105){\makebox(0,0)[r]{\strut{}$5$}}%
      \csname LTb\endcsname%
      \put(862,2492){\makebox(0,0)[r]{\strut{}$10$}}%
      \csname LTb\endcsname%
      \put(862,2879){\makebox(0,0)[r]{\strut{}$15$}}%
      \csname LTb\endcsname%
      \put(1285,372){\makebox(0,0){\strut{}$-20$}}%
      \csname LTb\endcsname%
      \put(1928,372){\makebox(0,0){\strut{}$-10$}}%
      \csname LTb\endcsname%
      \put(2570,372){\makebox(0,0){\strut{}$0$}}%
      \csname LTb\endcsname%
      \put(3213,372){\makebox(0,0){\strut{}$10$}}%
      \csname LTb\endcsname%
      \put(3855,372){\makebox(0,0){\strut{}$20$}}%
      \csname LTb\endcsname%
      \put(4498,372){\makebox(0,0){\strut{}$30$}}%
      \csname LTb\endcsname%
      \put(1157,2647){\makebox(0,0)[l]{\strut{}Linear model}}%
    }%
    \gplgaddtomacro\gplfronttext{%
      \csname LTb\endcsname%
      \put(361,1718){\rotatebox{-270}{\makebox(0,0){\strut{}Residual ($^\circ$)}}}%
      \csname LTb\endcsname%
      \put(2891,93){\makebox(0,0){\strut{}Rotation phase ($^\circ$)}}%
    }%
    \gplgaddtomacro\gplbacktext{%
      \csname LTb\endcsname%
      \put(5141,372){\makebox(0,0){\strut{}$-20$}}%
      \csname LTb\endcsname%
      \put(5784,372){\makebox(0,0){\strut{}$-10$}}%
      \csname LTb\endcsname%
      \put(6426,372){\makebox(0,0){\strut{}$0$}}%
      \csname LTb\endcsname%
      \put(7069,372){\makebox(0,0){\strut{}$10$}}%
      \csname LTb\endcsname%
      \put(7711,372){\makebox(0,0){\strut{}$20$}}%
      \csname LTb\endcsname%
      \put(8354,372){\makebox(0,0){\strut{}$30$}}%
      \csname LTb\endcsname%
      \put(5013,2647){\makebox(0,0)[l]{\strut{}Quadratic model}}%
    }%
    \gplgaddtomacro\gplfronttext{%
      \csname LTb\endcsname%
      \put(6747,93){\makebox(0,0){\strut{}Rotation phase ($^\circ$)}}%
      \csname LTb\endcsname%
      \put(9065,557){\makebox(0,0)[l]{\strut{}$0.01$}}%
      \csname LTb\endcsname%
      \put(9065,1494){\makebox(0,0)[l]{\strut{}$0.1$}}%
      \csname LTb\endcsname%
      \put(9065,2431){\makebox(0,0)[l]{\strut{}$1$}}%
      \csname LTb\endcsname%
      \put(9524,1718){\rotatebox{-270}{\makebox(0,0){\strut{}Subpulse peak flux (arb. units)}}}%
    }%
    \gplbacktext
    \put(0,0){\includegraphics{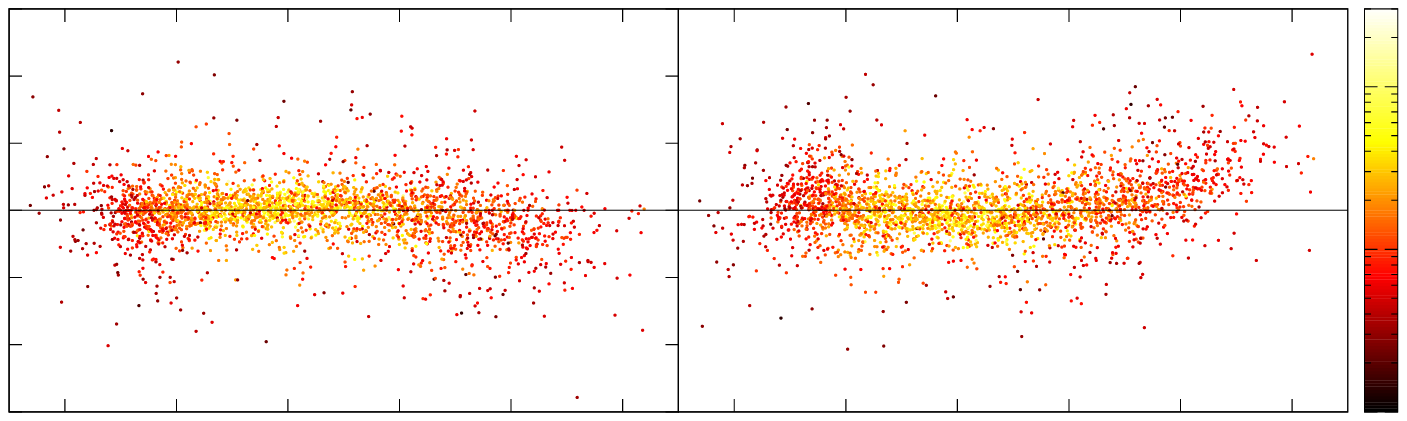}}%
    \gplfronttext
  \end{picture}%
\endgroup

%% file: P2s_vs_phase.tex
% GNUPLOT: LaTeX picture with Postscript
\begingroup
  \makeatletter
  \providecommand\color[2][]{%
    \GenericError{(gnuplot) \space\space\space\@spaces}{%
      Package color not loaded in conjunction with
      terminal option `colourtext'%
    }{See the gnuplot documentation for explanation.%
    }{Either use 'blacktext' in gnuplot or load the package
      color.sty in LaTeX.}%
    \renewcommand\color[2][]{}%
  }%
  \providecommand\includegraphics[2][]{%
    \GenericError{(gnuplot) \space\space\space\@spaces}{%
      Package graphicx or graphics not loaded%
    }{See the gnuplot documentation for explanation.%
    }{The gnuplot epslatex terminal needs graphicx.sty or graphics.sty.}%
    \renewcommand\includegraphics[2][]{}%
  }%
  \providecommand\rotatebox[2]{#2}%
  \@ifundefined{ifGPcolor}{%
    \newif\ifGPcolor
    \GPcolortrue
  }{}%
  \@ifundefined{ifGPblacktext}{%
    \newif\ifGPblacktext
    \GPblacktexttrue
  }{}%
  % define a \g@addto@macro without @ in the name:
  \let\gplgaddtomacro\g@addto@macro
  % define empty templates for all commands taking text:
  \gdef\gplbacktext{}%
  \gdef\gplfronttext{}%
  \makeatother
  \ifGPblacktext
    % no textcolor at all
    \def\colorrgb#1{}%
    \def\colorgray#1{}%
  \else
    % gray or color?
    \ifGPcolor
      \def\colorrgb#1{\color[rgb]{#1}}%
      \def\colorgray#1{\color[gray]{#1}}%
      \expandafter\def\csname LTw\endcsname{\color{white}}%
      \expandafter\def\csname LTb\endcsname{\color{black}}%
      \expandafter\def\csname LTa\endcsname{\color{black}}%
      \expandafter\def\csname LT0\endcsname{\color[rgb]{1,0,0}}%
      \expandafter\def\csname LT1\endcsname{\color[rgb]{0,1,0}}%
      \expandafter\def\csname LT2\endcsname{\color[rgb]{0,0,1}}%
      \expandafter\def\csname LT3\endcsname{\color[rgb]{1,0,1}}%
      \expandafter\def\csname LT4\endcsname{\color[rgb]{0,1,1}}%
      \expandafter\def\csname LT5\endcsname{\color[rgb]{1,1,0}}%
      \expandafter\def\csname LT6\endcsname{\color[rgb]{0,0,0}}%
      \expandafter\def\csname LT7\endcsname{\color[rgb]{1,0.3,0}}%
      \expandafter\def\csname LT8\endcsname{\color[rgb]{0.5,0.5,0.5}}%
    \else
      % gray
      \def\colorrgb#1{\color{black}}%
      \def\colorgray#1{\color[gray]{#1}}%
      \expandafter\def\csname LTw\endcsname{\color{white}}%
      \expandafter\def\csname LTb\endcsname{\color{black}}%
      \expandafter\def\csname LTa\endcsname{\color{black}}%
      \expandafter\def\csname LT0\endcsname{\color{black}}%
      \expandafter\def\csname LT1\endcsname{\color{black}}%
      \expandafter\def\csname LT2\endcsname{\color{black}}%
      \expandafter\def\csname LT3\endcsname{\color{black}}%
      \expandafter\def\csname LT4\endcsname{\color{black}}%
      \expandafter\def\csname LT5\endcsname{\color{black}}%
      \expandafter\def\csname LT6\endcsname{\color{black}}%
      \expandafter\def\csname LT7\endcsname{\color{black}}%
      \expandafter\def\csname LT8\endcsname{\color{black}}%
    \fi
  \fi
    \setlength{\unitlength}{0.0500bp}%
    \ifx\gptboxheight\undefined%
      \newlength{\gptboxheight}%
      \newlength{\gptboxwidth}%
      \newsavebox{\gptboxtext}%
    \fi%
    \setlength{\fboxrule}{0.5pt}%
    \setlength{\fboxsep}{1pt}%
\begin{picture}(9640.00,2880.00)%
    \gplgaddtomacro\gplbacktext{%
      \csname LTb\endcsname%
      \put(669,558){\makebox(0,0)[r]{\strut{}$0$}}%
      \csname LTb\endcsname%
      \put(669,1022){\makebox(0,0)[r]{\strut{}$10$}}%
      \csname LTb\endcsname%
      \put(669,1486){\makebox(0,0)[r]{\strut{}$20$}}%
      \csname LTb\endcsname%
      \put(669,1951){\makebox(0,0)[r]{\strut{}$30$}}%
      \csname LTb\endcsname%
      \put(669,2415){\makebox(0,0)[r]{\strut{}$40$}}%
      \csname LTb\endcsname%
      \put(669,2879){\makebox(0,0)[r]{\strut{}$50$}}%
      \csname LTb\endcsname%
      \put(1214,372){\makebox(0,0){\strut{}$-20$}}%
      \csname LTb\endcsname%
      \put(2101,372){\makebox(0,0){\strut{}$-10$}}%
      \csname LTb\endcsname%
      \put(2988,372){\makebox(0,0){\strut{}$0$}}%
      \csname LTb\endcsname%
      \put(3875,372){\makebox(0,0){\strut{}$10$}}%
      \csname LTb\endcsname%
      \put(4762,372){\makebox(0,0){\strut{}$20$}}%
      \csname LTb\endcsname%
      \put(4850,2647){\makebox(0,0)[l]{\strut{}(a)}}%
    }%
    \gplgaddtomacro\gplfronttext{%
      \csname LTb\endcsname%
      \put(270,1718){\rotatebox{-270}{\makebox(0,0){\strut{}$P_2$ ($^\circ$)}}}%
      \csname LTb\endcsname%
      \put(2988,93){\makebox(0,0){\strut{}Mean subpulse phase ($^\circ$)}}%
      \csname LTb\endcsname%
      \put(1415,2693){\makebox(0,0){\footnotesize Mode A\hspace{15pt}}}%
      \csname LTb\endcsname%
      \put(1415,2507){\makebox(0,0){\footnotesize Mode B\hspace{15pt}}}%
      \csname LTb\endcsname%
      \put(1415,2321){\makebox(0,0){\footnotesize Mode C\hspace{15pt}}}%
      \csname LTb\endcsname%
      \put(1415,2135){\makebox(0,0){\footnotesize Unknown\hspace{15pt}}}%
    }%
    \gplgaddtomacro\gplbacktext{%
      \csname LTb\endcsname%
      \put(5103,558){\makebox(0,0)[r]{\strut{}}}%
      \csname LTb\endcsname%
      \put(5103,1022){\makebox(0,0)[r]{\strut{}}}%
      \csname LTb\endcsname%
      \put(5103,1486){\makebox(0,0)[r]{\strut{}}}%
      \csname LTb\endcsname%
      \put(5103,1951){\makebox(0,0)[r]{\strut{}}}%
      \csname LTb\endcsname%
      \put(5103,2415){\makebox(0,0)[r]{\strut{}}}%
      \csname LTb\endcsname%
      \put(5103,2879){\makebox(0,0)[r]{\strut{}}}%
      \csname LTb\endcsname%
      \put(5629,372){\makebox(0,0){\strut{}$-20$}}%
      \csname LTb\endcsname%
      \put(6477,372){\makebox(0,0){\strut{}$-10$}}%
      \csname LTb\endcsname%
      \put(7326,372){\makebox(0,0){\strut{}$0$}}%
      \csname LTb\endcsname%
      \put(8174,372){\makebox(0,0){\strut{}$10$}}%
      \csname LTb\endcsname%
      \put(9022,372){\makebox(0,0){\strut{}$20$}}%
      \csname LTb\endcsname%
      \put(9107,2647){\makebox(0,0)[l]{\strut{}(b)}}%
    }%
    \gplgaddtomacro\gplfronttext{%
      \csname LTb\endcsname%
      \put(7325,93){\makebox(0,0){\strut{}Rotation phase ($^\circ$)}}%
    }%
    \gplbacktext
    \put(0,0){\includegraphics{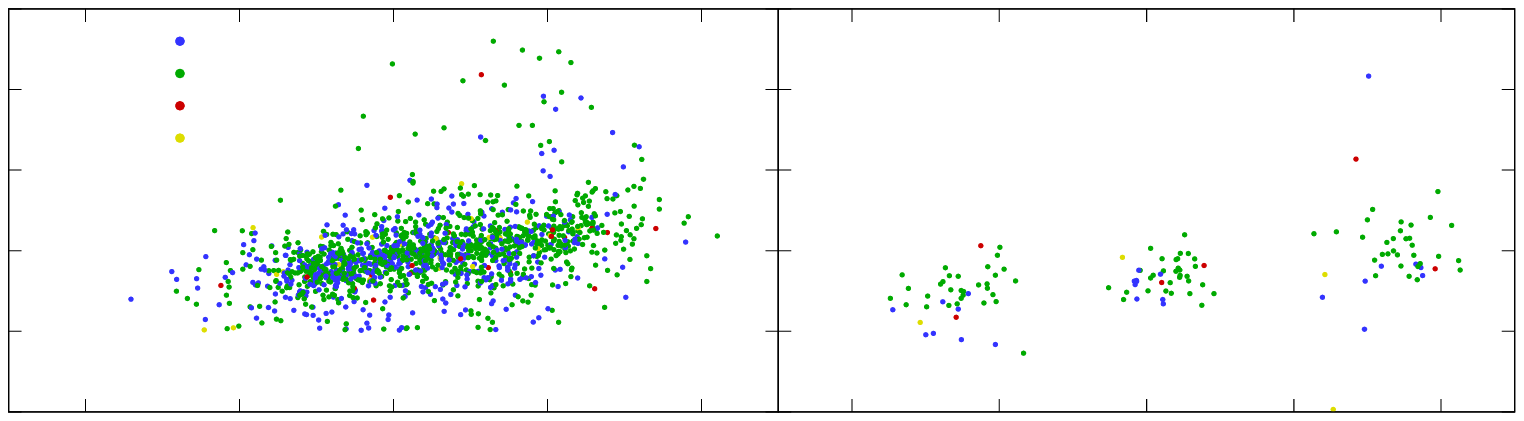}}%
    \gplfronttext
  \end{picture}%
\endgroup